\pdfoutput=1 
\documentclass[journal,12pt,onecolumn, draftclsnofoot]{IEEEtran}

\IEEEoverridecommandlockouts                              % This command is only
\usepackage{amsmath} % assumes amsmath package installed
\usepackage{pgf}
\usepackage{caption}
\usepackage[caption=false]{subfig}
\usepackage{graphicx} 
\usepackage[export]{adjustbox}
\usepackage{algorithm}
\usepackage{algorithmicx,algpseudocode}
\usepackage[normalem]{ulem}
\usepackage{lipsum}
\usepackage{mathtools}
\usepackage{cuted}
\usepackage{multirow}
\usepackage{graphicx}
\usepackage{epsfig}
\usepackage{cite}
\usepackage{booktabs}
\usepackage{multirow}
\usepackage{algorithm}
\usepackage{color}
\usepackage{bm}
\usepackage[letterpaper,top=0.75in,bottom=1in,left=1.03in,right=1.01in]{geometry}
\newcommand{\comment}[1]{}

\onecolumn

\title{\LARGE \bf
Dynamic Control of Functional Splits for Energy Harvesting Virtual Small Cells: a Distributed Reinforcement Learning Approach 
}

\author{\IEEEauthorblockN{Dagnachew Azene T., Marco Miozzo, Paolo Dini}\\
    \IEEEauthorblockA{CTTC/CERCA, Av. Carl Friedrich Gauss, 7, 08860, Castelldefels, Barcelona, Spain
    \\\{dtemesgene, mmiozzo, pdini\}@cttc.es}
    }
   
\begin{document}

\maketitle
\thispagestyle{empty}
\pagestyle{empty}

%%%%%%%%%%%%%%%%%%%%%%%%%%%%%%%%%%%%%%%%%%%%%%%%%%%%%%%%%%%%%%%%%%%%%%%%%%%%%%%%
\begin{abstract}
To meet the growing mobile data traffic demand, Mobile Network Operators (MNOs) are deploying dense infrastructures of small base stations as a solution for capacity enhancement. With densification, the power consumption of mobile networks and their impact on the environment are rising. As a result, we have seen a recent trend of powering base stations with ambient energy sources to achieve both environmental sustainability and cost reductions. In addition, flexible functional split in Cloud Radio Access Network (CRAN) is a promising solution to overcome the capacity and latency challenges in the fronthaul.  In such architecture, local base stations perform partial baseband processing while the remaining part will take place at the central cloud. As the base stations become smaller and deployed in densified manner, it is evident that baseband processing power consumption has a huge share in the total base station power consumption breakdown.  
In this paper, we propose a network scenario where the baseband processes of the virtual small cells powered solely by energy harvesters and batteries can be opportunistically executed in a grid-connected edge computing server, co-located at the macro base station site. We state the corresponding energy minimization problem and propose multi-agent Reinforcement Learning (RL) to solve it. Distributed Fuzzy Q-Learning and Q-Learning  on-line algorithms are tailored for our purposes. Coordination among the multiple agents is achieved by broadcasting system level information to the independent learners.  
The evaluation of the network performance confirms that coordination via broadcasting may achieve higher system level gains than un-coordinated solutions and cumulative rewards closer to the off-line bounds. Finally, our analysis permits to evaluate the benefits of continuous state/action representation for the learning algorithms in terms of faster convergence, higher cumulative reward and more adaptation to changing environments.

\end{abstract}

\begin{IEEEkeywords} 
Energy harvesting, Virtual small cells, Functional splits, CRAN, Reinforcement learning, Fuzzy Q-learning, Q-learning, Edge computing
\end{IEEEkeywords}

%%%%%%%%%%%%%%%%%%%%%%%%%%%%%%%%%%%%%%%%%%%%%%%%%%%%%%%%%%%%%%%%%%%%%%%%%%%%%%%%
\section{Introduction}
It is evident that there is an exponential growth of mobile traffic demand \cite{cisco}. 
To cope with this, Mobile Network Operators (MNOs) are deploying dense networks, which are composed of multi-tier Base Stations (BSs) in the same coverage area. This involves a network setup of many small BSs (SBSs) for satisfying the traffic demand from hot-spots (e.g. shopping malls, offices, entertainment areas) and Macro Base Stations (MBSs) to ensure mobility and reliable coverage~\cite{ultradense}. The resulting mobile network architecture is known as Heterogeneous Networks (HetNets). 
%The deployment of small base stations allow MNOs to increase their capacity in line with the growing traffic demand.
On the other hand, as mobile networks become densified, their electrical power requirements are also rapidly increasing. As a result, power consumption is playing a major part in the operational expenditures of MNOs. Moreover, there is an increasing concern regarding the environmental impact of Information and Communication Technology (ICT). It is estimated that ICT is consuming about $10\%$ of worldwide electricity generation and is forecasted that it might reach $53\%$ in a decade \cite{2015global}. Hence, energy sustainability is identified as one of the key requirements in the design and operation of mobile networks in order to ensure cost effectiveness and reduce the impact on the environment.

In the last years, Cloud Radio Access Network (CRAN) architecture has been proposed for enabling an efficient resource utilization via centralized processing of Baseband (BB) functions~\cite{chinamobile}. In CRAN, BB processing takes place at a centralized Baseband Unit (BBU) pool. As a result, base stations are reduced to simple Radio Remote Heads (RRHs). The connection between the RRHs and BBU pool is provided by the fronthaul links. CRAN ensures simplified base stations at cell sites and more efficient resource utilization due to centralized processing. The main drawback of CRAN is the need for a high capacity and very low latency fronthaul. As a result of efforts to relax the fronthaul requirements, flexible functional split between local BS sites and a central BBU pool is proposed~\cite{scforum}. Hence, part of the BB processes are executed at the local BS sites, while maintaining many of the centralization advantages of CRAN. Moreover, with the advent of Network Function Virtualization (NFV), network functions can be executed on general purpose computing hardware as virtual functions with Software Defined Networking (SDN) applied as a tool to realize the management and control of such functions~\cite{sdn-nfv}. As a result of Flexible Functional Splits and leveraging on NFV and SDN, network functions of SBSs, (e.g. BB functions), can be virtualized and placed at different sites of the network. These small base stations are known as Virtual Small Cells (vSCs) and enable flexibility in resource allocation and management. More recently, Multi-Access Edge Computing (MEC) has been introduced to enable the convergence of IT and telecommunication networking \cite{mec}. Thanks to MEC, BSs can leverage cloud computing capabilities and share part of their computational processes. % (e.g. baseband processing units).

As a means of ensuring energy sustainability, Energy Harvesting (EH) technology is becoming widely applicable in mobile networks. EH allows both cost and environmental impact reductions~\cite{piro2013}. However, EH comes with its own unique challenge mainly due to intermittent energy sources which cause unreliable supplying. Hence, in Energy Harvesting Base Stations (EHBSs), it is important to intelligently manage the harvested energy and to ensure proper energy storage provision to avoid outage. Consequently,  MEC enabled EHBSs can open a new frontier in energy-aware processing and sharing of BB processing units according to Flexible Functional Split options. The vSCs that completely rely on EH can opportunistically use the BB processing units available at the MEC server, which can be co-located at the MBS site. This is particularly important since the power consumption due to BB processing has a huge share in the total power consumption breakdown of smaller base stations. Consequently, MEC-enabled energy-aware placement of BB processes according to functional split options is a promising technique to enable energy efficient operation of vSCs powered by EH.

Overcoming the challenges that arise from EH and ensuring, an intelligent energy management scheme requires the design of dedicated control methods. In our previous work \cite{fss-vtc}, we have studied the performance bounds of dynamic placement of different functional split options 
%, namely PHY-RF, UpperPHY-LowerPHY, MAC-PHY, 
in an off-line manner for vSCs powered by EH. These performance bounds are determined by solving a joint grid energy consumption and system outage minimization problem, based on a-priori knowledge of the system dynamics (traffic and energy arrivals) subject to battery constraints. These results proof that dynamically adapting functional split options can provide significant grid energy saving as opposed to static configuration options. However, the off-line solution relies on a-priori knowledge and has limitation to scale up with the number of vSCs due to high computational complexity. On the other hand, Machine Learning (ML) tools can be used to extract models that reflect the user, energy harvesting and network behaviors, and to solve interactive decision making problems in real-time, at short time scales and with minimum a-priori information of the system \cite{knowledge}.
In particular, Reinforcement Learning (RL) based algorithm for dynamic placement of functional split options is proposed in our previous work \cite{dynamic-pimrc}. It is based on Temporal Difference (TD) learning methods, namely Q-learning and SARSA~\cite{suttonRL}, for on-line learning of control policies of a vSC powered by EH with flexible operative modes. The control has been implemented with an agent placed in the vSC and we investigated results for the case of single and autonomous vSC deployment~\cite{dynamic-pimrc}. In this case, RL allows learning of optimal strategy through interaction with the system environment for achieving system wide objectives, i.e. efficient utilization of the harvested energy. However, when considering the case of various vSCs operating simultaneously, RL is expected to face more problems. Centralized solutions might experience long convergence and training phases due to the high number of state/action pairs needed to model the environment. 
Alternatively, a distributed approach may allow to reduce the complexity by dividing the problem among the multiple agents. Nevertheless, multi-agent systems can experience issues due to the conflicting interests of the agents \cite{marl-survey}.
%Alternatively, a distributed approach, also called multi-agent, allows to reduce the complexity by dividing the problem among the agents. Nevertheless, multi-agents systems can experience issues due to the conflicting interests of the agents. 
In fact, when each vSC is allowed to learn the best energy management policy independently, there is a risk that its actions affect other vSCs' policies, which in turn would have a negative effect on the overall performance of the network, e.g. system drop rate. Hence, multi-agent RL based strategy should ensure coordination among the learning agents, i.e. the vSCs, towards achieving system wide gains. This paper proposes multi-agent RL based on-line algorithms for dynamic placement of functional split options in MEC-enabled RAN with energy harvesting capabilities. Both Fuzzy Q-Learning (FQL) and Q-Learning (QL) based on-line algorithms are proposed with performance comparisons and evaluation against an off-line bound. Coordination among the multiple agents is achieved by broadcasting system level information to the independent learners. A comparison with an implementation of the learning algorithms without coordination is also analyzed. 

The main contributions of the paper may be summarized in the following items:

\begin{itemize}
	                   
	\item \textit{Edge Computing Platforms for Energy Saving}: we propose a network scenario where part of the computational processes of vSCs powered solely by energy harvesters may be shared with on grid-connected central MEC-server at the MBS. 
	\item \textit{Grid Energy Minimization Problem Statement}: we formulate a network wide grid energy optimization problem while avoiding system outage for the proposed network scenario. 
	%multi-vSCs that are fully powered by EH plus batteries and rely on grid-connected central MEC-server at the MBS for part of their BB processing
	\item \textit{Coordinated Multi-agent RL Solutions}: we propose multi-agent RL controllers to solve the grid energy optimization problem. In particular, distributed FQL and QL based solutions are tailored for our purposes, including different levels of coordination among the vSCs.
    \item \textit{Characterization of the Learning Algorithms}: we analyze the complexity and the convergence of the proposed learning algorithms by giving insights of the hyperparameter setup in both simulative training and run-time scenarios. We study the effects of the quantization of states and actions on the stability and system performance. We characterize the selected policy of the coordinated solutions with respect to the off-line bound. 
    %Simulated training of the proposed FQL and QL control methods including analysis on the choice of training parameters.
	\item \textit{Network Performance Evaluation}: we evaluate the network performance (in terms of energy consumption and traffic drop rate) by our multi-agent RL solutions with different levels of coordination and compare them against the off-line performance bound and static solutions. 
	%evaluation of network performances of the trained control methods with respect to the off-line performance bound and benchmark RL solutions without co-ordinations. Moreover, validation of the trained FQL and QL methods are also included.
\end{itemize}

We state here that the list of contributions of our paper represents the first effort on controlling MEC-enabled RAN with energy harvesting capabilities through coordinated multi-agent RL. For more details, the reader is referred to Section II, in which we survey the related work on the topic.

The rest of the paper is organized as follows. Section~\ref{sec:relatedwork} describes the related literature and Section~\ref{sec:architecture} shows the reference architecture considered in this work. Section~\ref{sec:system-model} describes the overall system model including power consumption, network, traffic and energy arrival models. Both FQL and QL based control designs are explained in~\ref{sec:control}.  Section~\ref{sec:results} shows the simulation scenario and numerical results of simulations including the comparison among QL, FQL and off-line solutions. Finally, we draw our conclusions in Section~\ref{sec:conclusions}.

\section{Related work}
\label{sec:relatedwork}

Recently, intelligent energy management in EHBSs has been the focus of many studies in the research community due to the increasing significance of energy sustainability combined with dense deployment of BSs. Most of these literature analyze hierarchical multi-tier networks, the so called HetNets, with an intelligent switching on/off scheduling of BSs. The authors in~\cite{piovesan2017} apply Dynamic Programming (DP) to determine the optimal switch on/off policy in a two-tier HetNets with baseline MBS and hot-spot deployed SBSs. The solution shows the performance bound of an intelligent switch on/off policy when all the system dynamics information are known a-priori. Minimizing the grid energy consumption for hybrid powered base stations is also studied in \cite{gong2014}. Here, the authors applied a two stage DP methods designed to achieve energy saving gains while maintaining probability of blocking. The authors in~\cite{lee2017} apply a ski-rental framework based on-line algorithm for optimal switch on/off scheduling for minimizing the operational costs of a network composed of self powered base stations. Moreover, the authors in \cite{zhou2013} studied sleep mode coordination between base stations powered by EH and grid energy using DP. However, the DP based solution is shown to entail high computational complexity. On the other hand, the authors in~\cite{miozzo2015} apply RL, in particular QL algorithm, to optimize the harvested energy utilization. The work is based on distributed Q-learning where each renewable powered base station take autonomous decision whether to switch on/off according to energy arrival, energy storage and traffic demand. In addition, multi-armed bandit based distributed learning is studied in \cite{ameur2016} to allow each SBS to learn its own energy-efficient policy. 
%The authors in \cite{lin2015} proposed RL based energy cooperation among nodes that are capable of EH and are able to transmit harvested energy to other nodes \emph{[MM: this is on cooperation, can be removed]}. 
The authors in \cite{ll-vtc} applied layered learning for system wide harvested energy allocation through decomposition of the problem into two layers. The first layer, based on RL, is in charge of local control at each SBS and the second layer, based on artificial neural networks, ensures network wide coordination among the SBSs. Moreover, renewable energy allocation in edge computing devices with EH is studied in \cite{xu2016}. Here, the authors propose RL based on-line solutions for offloading and auto-scaling in edge computing devices that are powered by EH. On the other hand, the authors in \cite{mendil} proposed a RL based energy controller for a SBS powered by energy harvesting, battery and smart grid by considering battery aging effects. This work is based on FQL and is shown to provide significant extension to the life time of a small cell battery. However, this work is limited to a single SBS and a coordinated energy management among base stations with in a mobile network remains an open issue. 

Nevertheless, there is a gap in the literature in integrating EH and flexible functional split options in MEC-enabled RAN: to the best of our knowledge, no solution has analyzed the possibility of dynamically sharing the BB processes of vSCs with a MEC server co-located with the MBS, which is the main topic of our paper. Here, we claim that functional splits give insight into considering more operative modes of small BSs, in addition to switch on and off and enable higher grid energy savings. Moreover, from a methodological perspective, most of the work has used RL to solve a single agent problem. Instead, in this paper, we propose a multi-agent solution for a multi-cell scenario using multi-agent RL algorithms with different levels of coordination among agents. In particular, we propose to communicate system level information to the multiple agents and compare such coordination with solutions based on independent learners. Moreover, we tailor QL and FQL algorithms to our network scenario and evaluate the effects of the quantization of the states in the system performance.

%The functional splits give insight into considering more operative modes of BSs, in addition to switch on and off. Hence, this paper focuses on designing both QL and FQL based intelligent energy controls in mobile networks with multi-vSCs deployments that rely on a central MBS and MEC server for performing part of their BB processing.

\section{Reference Architecture}
\label{sec:architecture}

\begin{figure}
	\centering
	\includegraphics[scale=1.8]{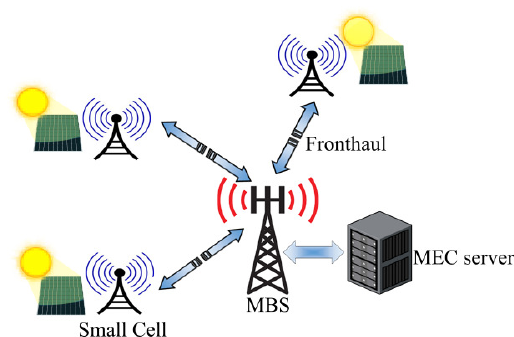}
	\caption{Reference architecture}
	\label{fig:scenario}
\end{figure}

This work considers a two tier network architecture. The first tier consists of a MBS and co-located BBU pool. The central BBU pool is hosted at the MEC server located at MBS site. MBS with MEC server are responsible for providing baseline coverage, mobility support and baseband processing resources. The MBS site is fully powered by energy from the grid, thus assuring reliable communications and computing. The second tier is composed of vSCs, which are deployed in hot-spot manner for capacity enhancement and they do not overlap in coverage \cite{ngmn}. They are completely powered by solar panel and batteries and they are fully or partially dependent on the MEC server at the MBS for BB processing. The reference architecture is shown in Fig. \ref{fig:scenario}.

The proposed MEC-enabled architecture jointly with SDN and NFV paradigms~\cite{nfv-cran} are enablers for automated network management, flexibility and cost reductions. In MEC deployments,  multi-tier MEC servers are co-located at the BSs and have different computational and transmission capabilities (i.e., MBSs are high-power and high-computing nodes) \cite{Park2018}. In this work, we are interested in the case that MBSs may support vSCs and enable computational offloading of some of their networking functions. Hence, the vSCs opportunistically use the central BBU pool, i.e. the MEC server at MBS, for full or partial BB processing requirements. For doing so, a standardized interface (e.g. Openflow~\cite{Lara2014}), can be used to implement the interactions between the MBSs and vSCs. In this way, vSCs transmission functions are decoupled from proprietary hardware-dependent implementations and may be executed in a different hardware resource of the network. To this respect, 3GPP has defined different functional splits between the distributed and the centralized unit~\cite{3gpp-38801}, in our case the vSCs and the MBS, respectively.
The vSCs in our scenario can opportunistically operate in one of the functional split configuration options, which are based on~\cite{3gpp-38801} and are explained below. 
\begin{itemize}
	\item PHY-RF split: all the protocols, Physical (PHY) and above layers, are implemented at the MEC server located in the MBS site. Hence, the vSC behaves as a Radio Frequency (RF) transceiver, used only for signal transmission and reception;
	\item MAC-PHY split: PHY layer processing takes place at the vSC, in addition to RF functions.  Medium Access Control (MAC) and above layer functions are executed by the MEC server at the MBS site. 
\end{itemize}

These two functional split options have been selected based on their impact on the energy consumption of vSCs. In fact, PHY-RF and MAC-PHY are the options that have larger variations in terms of energy expenditure, which allows to implement a dynamic control on them. A finer grade would not introduce more flexibility, since the others split options have a negligible impact on the energy utilization, as demonstrated in our previous work~\cite{fss-vtc}.

\begin{figure}
	\centering
	\includegraphics[scale=0.6]{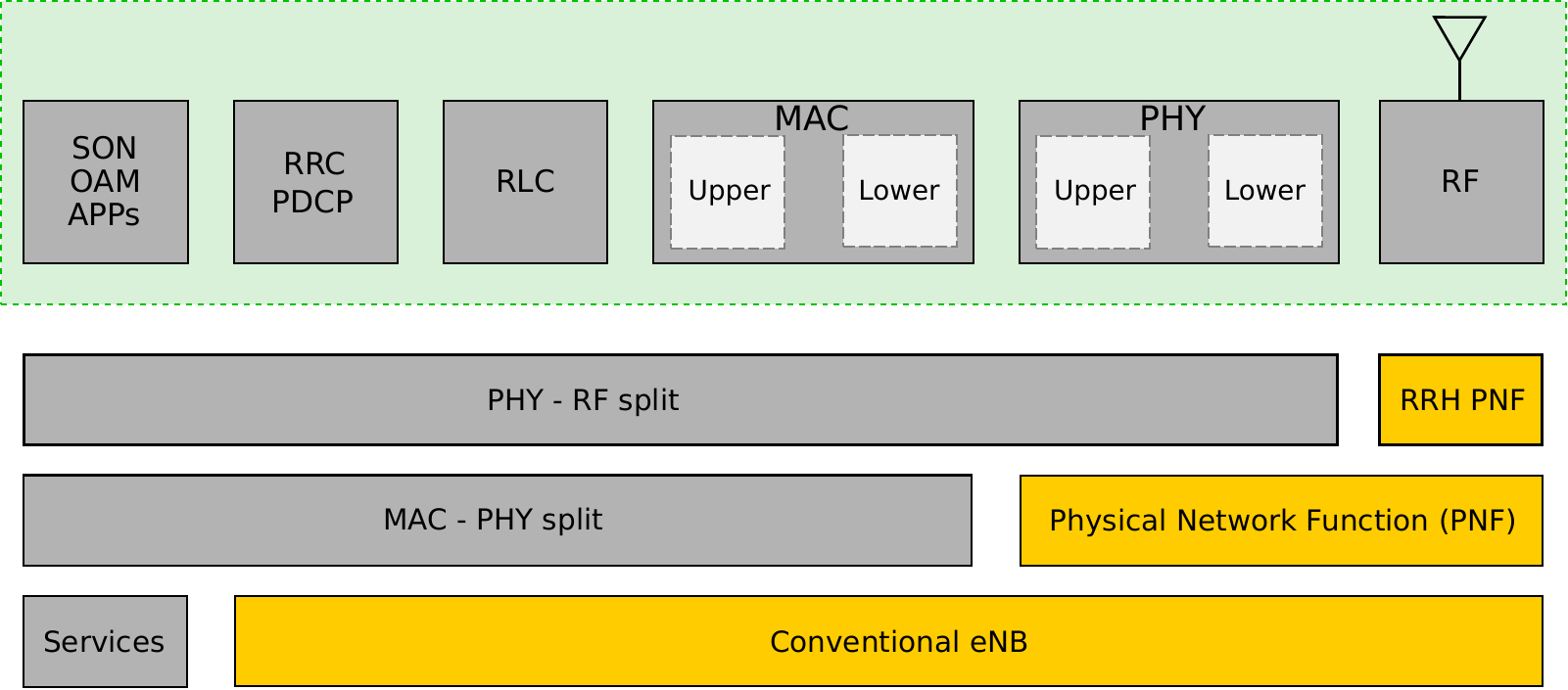}
	\caption{The different implementations of the functional split configuration options for a vSC, including PHY-RF and MAC-PHY split. The conventional eNodeB (eNB) configuration is also shown for comparison.}
	\label{fig:split}
\end{figure}

The functional split options are depicted in Fig.~\ref{fig:split} along with the conventional eNodeB architecture.
Each operative mode corresponds to different computational load for the vSCs and MBSs, which in turn, corresponds to different energy consumption models, as will be described in Section~\ref{bs-power-model}. It is worth highlighting here that in PHY-RF split, the vSCs are executing only the RF functionalities and the other upper layer functions, including PHY and MAC, are executed at the MBS site. This is similar to a Cloud RAN (C-RAN) architecture. Therefore, \mbox{MEC} deployment relies on a reconfigurable front-haul, since the bandwidth and latency requirements become more stringent when more functions are placed in the centralized unit~\cite{scforum}.\\

\section{ System Model}
\label{sec:system-model}

\subsection{Network Model}
\label{sec:network-model}
%\emph{(MM: Let me suggest you to had a look to the system model of the journal on LL, it would be better to explain here also why QL and FQL can approximate the problem in (2), e.g., presenting the general MDP problem)}

We consider a two-tier mobile network composed of clusters of one MBS with co-located BBU pool and $N$ vSCs.
The system evolves in time slots based on the variation of the traffic demand and the energy arrivals. The traffic load at slot $t$ generated by the users in the coverage of the vSCs is defined as $\boldsymbol{L}^t= [L_1^t,L_2^t,\ldots,L_N^t ]$. The traffic load experienced by the MBS in slot $t$ is defined as $\boldsymbol{\rho}^t = [\rho_1^t,  \rho_2^t,\ldots,\rho_N^t]$.  The energy harvested by the vSCs in slot $t$ is defined by $\boldsymbol{H}^t= [H_1^t,H_2^t,\ldots,H_N^t ]$ and the energy stored by each vSC in its battery in slot $t$ is defined by $\boldsymbol{B}^t= [B_1^t,B_2^t,\ldots,B_N^t ]$. 
Moreover, the energy stored in the batteries at the beginning of the next slot, $\boldsymbol{B}^{t+1}$, is evaluated according to the following formula:
\begin{equation}
\boldsymbol{B}^{t+1}=\min \left(\boldsymbol{B}^t+\boldsymbol{H}^t-\boldsymbol{P}^t\Delta_t,B_{\mathrm{cap}}\right)
\label{eq:battery}
\end{equation}
where $\boldsymbol{P}_t= [P_1^t,P_2^t,\ldots,P_N^t ]$ is the power consumed by the vSCs in slot $t$ (and described in Section~\ref{bs-power-model}), $B_{\mathrm{cap}}$ is the maximum battery capacity and  $\Delta_t$ is the time difference between two consecutive slots.
The operative state of the vSCs in slot $t$ is defined by $\boldsymbol{A}^t=[A_1^t,A_2^t,\ldots,A_N^t ]$. At each slot, intelligent decisions are made to determine the optimal configuration of the vSCs in the mobile cluster to serve the traffic demand based on their available energy budget, energy arrival information and the traffic request. The optimization problem is defined by a Markov Decision Process (MDP) as $\boldsymbol{X}_{t+1}=f(\boldsymbol{X}_t,\boldsymbol{A}_t,\boldsymbol{w}_t)$, where $\boldsymbol{X}_t=[X_1^t,X_2^t,\ldots,X_N^t ]$ is the state of the vSCs in slot $t$, $\boldsymbol{A}_t=[A_1^t,A_2^t,\ldots,A_N^t ]$ is the control action and $\boldsymbol{w}_t=[w_1^t,w_2^t,\ldots,w_N^t ]$ is the random disturbance of the environmental variables. In particular, we define each state $X_i^t$, with $i=1,...,N$, as $X_i^t = (H^t_i, B^t_i, L^t_i, \rho^t_i)$ and the control action $A_i^t$ as follows:
\begin{equation}
A^t_i=
\begin{cases}
0 \quad \textrm{if the } i \textrm{-th vSC is OFF}\\
1 \quad \textrm{if the } i \textrm{-th vSC is in PHY-RF split mode}\\
2 \quad \textrm{if the } i \textrm{-th vSC is in MAC-PHY split mode}\\
\end{cases}
\end{equation}

The optimization objective is to minimize the energy consumption from the grid while avoiding system outage. We define system outage as the event to not satisfy the traffic demand due to battery energy depletion or wrong configuration decisions, which may overload the MBS with the traffic of the switched off vSCs. Hence, the general optimization problem can be formulated as follows: 
\begin{equation}
\begin{split}
\text{\boldmath{P1:}}\min_{\{\boldsymbol{A}^t\}_{t=1,\dots,K}}\; &\sum_{t=1}^K f(\boldsymbol{A}^t,t) \\
\text{subject to  } B_i^{t}&>B_{\mathrm{th}} \; \forall i .\\
\\
\end{split}
\label{eq:optimization}
\end{equation}
where $B_{\mathrm{th}}$ is the battery threshold level and $K$ is the time horizon or the number of times the energy control is applied; $f(\boldsymbol{A}^t,t)$  is the weighted cost function in slot $t$, which is defined as:
%\emph{[MM: is this f the same of (4)]}
\begin{equation}
f(\boldsymbol{A}^t,t)=\omega_1\cdot \mathrm{E_m}(\boldsymbol{A}^t,t)+ \omega_2\cdot {\mathrm{D}}(\boldsymbol{A}^t,t)
\label{eq:cost}
\end{equation}
where $\mathrm{E_m}(\boldsymbol{A}^t,t)$ and  $\mathrm{D}(\boldsymbol{A}^t,t)$ are respectively the normalized grid energy consumption and the traffic drop rate in the cluster, given the operative modes of the vSCs in slot $t$. 
The grid energy consumption in the slot~$t$ is equivalent to the energy consumption at the MBS site, including the MEC server used for computational offloading. The  grid energy consumption is then computed as:
\begin{equation}
%^\mathrm{E_m}(\boldsymbol{A}^t,t)= \frac{\mathrm{P_m}(\boldsymbol{A}^t,t)\Delta_t }{\mathrm{P_m^{MAX}}\Delta_t}
\mathrm{E_m}(\boldsymbol{A}^t,t)= \mathrm{P_m}(\boldsymbol{A}^t,t)\Delta_t 
\label{eq:mbs-energy}
\end{equation}
where $\mathrm{P_m}(\boldsymbol{A}^t,t)$ is the power consumption of the MBS given the operative modes of the vSCs.  
%whereas $\mathrm{P}_\mathrm{m}^\mathrm{MAX}$ is the power consumption of the MBS at full load including the energy required due to the BB processes from the vSCs. 
The details on the power consumption models are described in Section IV.B.
%(PD: I cannot understand such power model. Why the denominator is needed? For normalization purposes? If so, it is really needed to show it in a formula?)
%(\emph{MM: including the $P_{BB}$ of all the vSCs?)} 
%and $\Delta_t$ is the time difference between two consecutive slots. %In case the total excess energy shared by the SBSs is greater than the MBS power consumption a time $t$, the redundant energy is dissipated locally at the SBS site. 
%Moreover, we define system outage as the event to not satisfy the traffic demand due to battery energy depletion or wrong configuration decisions, which may overload the MBS. 
The traffic drop rate in slot $t$, $\mathrm{D}(\boldsymbol{S}^t,t)$, is the ratio of the total amount of traffic demand that cannot be served by the system in the slot $t$. Additionally, each battery at the vSCs has to be maintained in the proper State Of Charge (SOC) (i.e, above the battery level threshold $B_{\mathrm{th}}$) to avoid a rapid reduction of its lifetime~\cite{Lu2013}.
Finally, the weights $\omega_1$ and $\omega_2$  provide flexibility in the cost function to emphasize one part of the cost over the other. They must always be positive and sum to~1, that is, $\omega_1\geq 0$, $\omega_2\geq 0$, $\omega_1+ \omega_2=1$.  In this work, we will consider $\omega_1 = \omega_2 = 0.5$ to have a balanced importance of the two components. 

An off-line solution of this problem is proposed in our work~\cite{fss-vtc} using DP and with a priori knowledge of the environmental variables. The problem of finding optimal configuration options is represented as a graph and stated as a Shortest Path search, while Label Correcting Method is used to explore the graph and find the the shortest path.	
%In detail, the Graph Theory has been used with a Shortest Path search, which allows to find the ON/OFF using the Label Correcting Method to explore the graph. 
%However, the off-line solution is not scalable with the number of vSCs and relies on a-priori knowledge. Hence, 
Those obtained results are considered as system performance bounds and are used as a benchmark to the control methods proposed in this paper.

In this work, we propose an on-line solution based on multi-agent RL. In particular, we use approximated DP methods, known as TD learning, to determine optimal policies \cite{suttonRL}.
%are proposing temporal difference based RL solutions which are known as approximate DP methods to determine optimal policies \cite{suttonRL}. 
Our proposal is based on distributed and coordinated decision making: i.e. each vSCs take actions by them selves, which makes it scalable with the number of vSCs. In order to coordinate the decision making process, we rely on communicating system wide information, e.g. traffic load at MBS, to each vSCs. Section \ref{sec:control} describes the proposed QL and FQL based control solutions to the MDP described here.

\subsection{Power model}
\label{bs-power-model}
The power consumption of each split option is estimated based on the model introduced in \cite{desset2012}, which is a general flexible power model of LTE base stations and provides the power consumption in Giga Operation Per Second (GOPS). Technology dependent GOPS to Watt conversion factor is applied to determine the power consumption in Watts. 
In this paper, we have mapped the various BB processing tasks of the functional split options to their power requirement estimations. 
 %The fronthaul latency and bandwidth requirements, shown in Fig. \ref{fig:signalflow}, are estimated based on \cite{scforum}. %Therefore they are conservative for the scenario considered here, since they are not contemplated for HCRAN architecture. 

%\begin{figure}[b]
%\centering
%\includegraphics[scale = 0.1]{signalflow_2.eps}
%\caption{BBU functions with the considered functional splits and the relevant fronthaul latency and bandwidth requirements (estimations based on \cite{scforum})} 
%\label{fig:signalflow}
%\end{figure}

The total BS power consumption is given by:
\begin{equation}
P_{\mathrm{BS}}=P_{\mathrm{BB}}+P_{\mathrm{RF}}+P_{\mathrm{PA}} + P_{\mathrm{overhead}}
\label{eq:Ptotal}
\end{equation}
where $P_{\mathrm{BB}}$ is the power consumption due to the baseband processing, $P_{\mathrm{RF}}$ is the power consumption due to RF circuitry, $P_{\mathrm{PA}}$ is the power consumption by the power amplifier and $P_{\mathrm{overhead}}$ is the overhead power consumption (e.g., cooling system). 

Baseband power consumption, $P_{BB}$ is given by:
\begin{equation}
P_{\mathrm{BB}}  =[P_{\mathrm{CPU}}+ P_{\mathrm{OFDM}}+ P_{\mathrm{filter}} + P_{\mathrm{FD}} +  P_{\mathrm{FEC}}]
\label{eq:PBB}
\end{equation}
where $P_{\mathrm{CPU}}$ is the idle mode power consumption, $P_{\mathrm{OFDM}}$ is the power consumption due to OFDM processes, $P_{\mathrm{filter}}$ is the power consumption due to filtering,  $P_{\mathrm{FD}}$ is the frequency domain processing power consumption and $P_{\mathrm{FEC}}$ is the power consumption due to forward error correction (FEC) processes.
%In accordance with~\cite{desset2012}, only $P_{\mathrm{FD}}$  and $P_{\mathrm{FEC}}$  are dependent on the instantaneous traffic load. These power consumption values mainly depends on bandwidth, number of antennas and the load fraction. 
In accordance with~\cite{desset2012}, all the terms in (\ref{eq:PBB}) are dependent on the number of antennas and bandwidth. Moreover, $P_\mathrm{FD}$ and $P_\mathrm{FEC}$ are the only components that depend on the traffic load.

When the vSCs are in PHY-RF split mode, their power consumption model does not include the corresponding $P_{BB}$,  since the baseband processing takes place at the MBS site. Instead, the corresponding  $P_{BB}$ term is added to the MBS. On the other hand, in MAC-PHY split mode, the vSC power consumption includes the baseband power consumption term and is given in (\ref{eq:Ptotal}). % In fact, we claim here that the power consumption of a vSC in MAC-PHY split mode is almost equivalent to a conventional eNodeB due to the negligible impact of the MAC and upper layers implementation on the energy budget \emph{(MM: any ref?)}. 
 Considering the aformentioned model description, the grid power consumed by the MBS is computed as:

\begin{equation}
P_\mathrm{m} =  P_\mathrm{BS}^\mathrm{MBS} + \sum_{i \in \mathcal{G}}P_\mathrm{BB}^i
\label{eq:PBBsbs}
\end{equation}
where $P_\mathrm{BS}^\mathrm{MBS}$ is the power consumption of the MBS computed as in (\ref{eq:Ptotal}), $P_\mathrm{BB}^i$ is the baseband power consumption of the $i$-th vSC and $\mathcal{G}$ is the set containing the vSCs in PHY-RF split mode. 

\subsection{Energy Harvesting and Demand Profiles}
\label{sec-eh-traffic-profiles}
Hourly energy generation traces from a solar source have been obtained for the city of Los Angeles (CA, USA).
The solar raw irradiance data have been collected from the National Renewable Energy Laboratory and converted into harvested energy traces using the SolarStat tool~\cite{Miozzo2014}. %, accounting for a specific solar power technology,.
The energy harvesting traces are generally bell-shaped with a peak around midday, whereas the energy harvested during the night is negligible. Moreover, as discussed in~\cite{Miozzo2014}, high variability of the harvested energy may occur during the day and this also holds for the summer months. This means that, although the energy inflow pattern can be known to a certain extent, intelligent and adaptive algorithms that make their decisions based on current and past inflow patterns, as well as predictions of future energy arrivals, have to be designed. 

\begin{figure*}
\centering
\hspace*{-2cm}\includegraphics[scale = 0.6]{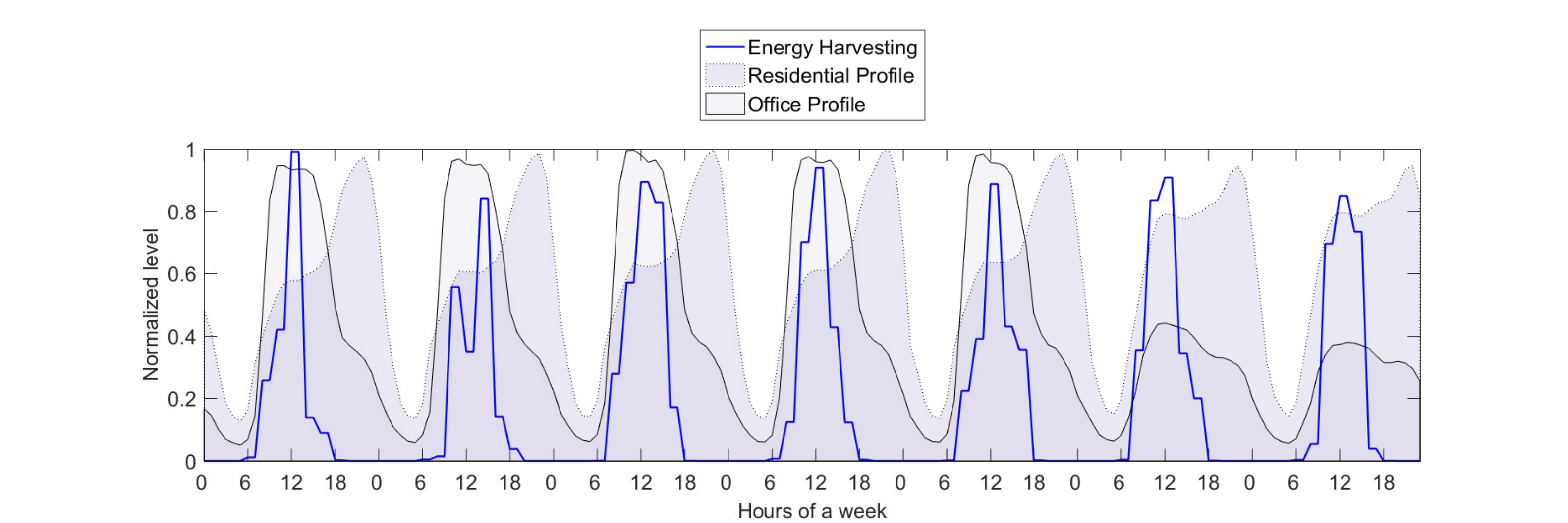}
\caption{Typical weekly normalized energy harvesting, office traffic profile and residential traffic profile } 
\label{fig:traffic-eh}
\end{figure*}

For the demand profile, it is commonly accepted and confirmed by measurements that the energy use of base stations is time-correlated and daily periodic. The UEs have been classified as in~\cite{earth-D23} in \emph{heavy} and \emph{ordinary} users according to their amount of requested traffic. Moreover, in this article we use the traffic load profile obtained in~\cite{Xu2017understanding} as the average amount generated by the users. In addition, based on the average traffic generated by the users, traffic variability is added following a normal distribution using standard deviation from measurements of real mobile traffic traces~\cite{mobiletraffic}.
%... (PD: Please add the way use implement traffic variability).
%(\emph{MM: What about the variance we add in deterministic traffic profile?}. 
The traffic demand of each UEs in a cycle are dimensioned based on traffic profiles presented in~\cite{Xu2017understanding}, which are derived from time, location and frequency information of thousands of cellular towers. The analysis in~\cite{Xu2017understanding} demonstrates that the urban mobile traffic usage can be described by mainly five basic time domain patterns that corresponds to different functional regions, i.e., residential, office, transportation, entertainment and comprehensive. In this article, we are considering residential and office profiles which are the most common use cases for urban deployment scenarios. An example of a normalized energy harvesting trace, the residential traffic profile and the office traffic profile for both week and weekend days is shown in Fig. \ref{fig:traffic-eh}. The figure shows that energy harvesting and residential traffic profile peaks occur at different hours of the day i.e. energy harvesting peak occurs around noon where as traffic demand peak occurs during the evening. This calls for an intelligent energy management to maximize the utilization of harvested energy. %\emph{(MM: Describe more accurately the work in~\cite{Xu2017understanding} and how it is applied to this work)}

\section{Distributed Control Methods}
\label{sec:control}

In this section, we introduce our distributed and coordinated multi-agent RL solutions and we focus on the details about both FQL and QL based controllers design. The section starts with background information on RL and Fuzzy Inference Systems (FIS).
Then, the two solutions based on independent learners without coordination are presented here.

\subsection{Background}

RL is a learning paradigm that relies on learning by interacting with the environment without an exemplary supervision \cite{suttonRL}. It is a well known framework of solving a problem described as an MDP. Formally, the RL framework is defined in terms of states, actions and rewards. Through the RL process, according to the current state, the agent executes a certain action and receives an immediate reward and as a result of the action, its environment will evolve to a new state. It is important to note that in RL, the rewards can be delayed. Hence, it is a sequential decision making process with the goal of maximizing cumulative reward. For our network model, the objective of the RL based controller is to learn energy management policies through the interaction with the environment. The controller decides the operative mode of vSCs in a cluster at each time slot %\emph{(MM: be careful, both time and time slot are used in this section, choose just one type, e.g., time-continuous, time slot-discrete, I would go with the discrete one)} 
based on the traffic load, energy arrival and energy storage information. 

Let $\boldsymbol{X}_t$ be the state of the system at time $t$, the controller chooses an action $A_t$ from action set $\boldsymbol{A}$, which translates to the operative modes of the vSC. As a result of this action, the environment returns an immediate reward $r_t$. Based on this $r_t$, the correspondent Q-value, $Q(\boldsymbol{X}_t,A_t)$, which represents the level of goodness when taking a specific action in a given state, will be updated. The process of learning needs a balance between \emph{exploration} i.e. taking random actions to discover new knowledge and \emph{exploitation} i.e. taking the actions that have been already discovered as good, i.e., the actions with the maximum Q-value. This process of selecting a specific action and updating the Q-value continues sequentially for each time slot. The controller selects the action at the beginning of each time slot $t$ based on the specific RL algorithm it applies. The goal of such algorithms is to determine iteratively the Q-values for each state-action pairs for achieving an optimal policy in the long-term.

%\emph{[MM: What about "distributed"? We need to introduce also the fact that we are considering a multi-agent system.]}\\
We propose a multi-agent RL solution based on a distributed control architecture. In fact, our reference network scenario consits of multiple vSCs in the coverage of a MBS. Therefore, single agent RL methods may have high complexity and slow convergence due to the high number of state/action pairs that represent the system. Instead, multi-agent RL methods guarantee higher scalability since they distribute the algorithms among the different vSCs. 
%The RL based control proposed in this section is based on a distributed approach. 
%Multiple agents, one per each vSCs, are in charge of learning the optimal strategy by dynamically interacting with the environment, an approach also known as multi-agent reinforcement learning.
%In such multi-agent systems, each agent (located at each vSC) locally learns part of a system wide optimal policy in an environment which is also affected by the actions of other agents. 
However, while in single-agent RL, the state of the environment changes solely as a result action by the agent, in multi-agent RL scenarios, the state of the environment is subject to actions from multiple agents. As a result, multi-agent RL is prone to conflicting interests among the learning agents and requires coordination techniques among the agents to learn optimal system-wide strategy. We solve the coordination issue by broadcasting system wide information to every vSCs, which harmonize the selected policies and achieve system wide gains.

On the other hand, FIS is the process of mapping a set of input control signals to a set of output actions through fuzzy rules \cite{tsoukalas1997fuzzy}. FIS are mainly applicable in systems that cannot be represented by explicit mathematical models through approximation of system knowledge in a similar way to human perception and reasoning.  The design of FIS involves the following steps \cite{tsoukalas1997fuzzy}:
\begin{enumerate}
	\item \textit{Fuzzification of the crisp input signals}: crisp values are the exact values as read from sensors or measurements. Fuzzification of crisp input signals is done by defining the fuzzy sets and membership functions of the input signals. Hence, the state space is partitioned into various fuzzy sets through membership functions. Each fuzzy set is associated with linguistic terms such as "high" or "low". The most common membership functions are triangular and trapezoidal.
	\item \textit{Defining the rule base}: this step involves defining the behavior of the controller in terms of control actions using linguistic variables defined in the first step. This corresponds to a series of "if ... then" rules for each combinations of fuzzy sets of input signals. For deriving these rules, expert knowledge and experience is generally used.
	\item \textit{Defuzzification}: this step reverts back the results from the fuzzy rule base in step $2$) back to crisp mode  and activates the output action.
\end{enumerate}
%1) \textit{Fuzzification of the crisp input signals}: crisp values are the exact values as read from sensors or measurements. Fuzzification of crisp input signals is done by defining the fuzzy sets and membership functions of the input signals. Hence, the state space is partitioned into various fuzzy sets through membership functions. Each fuzzy set is associated with linguistic terms such as "high" or "low". The most common membership functions are triangular and trapezoidal.2) \textit{Defining the rule base}: this step involves defining the behavior of the controller in terms of control actions using linguistic variables defined in the first step. This corresponds to a series of "if ... then" rules for each combinations of fuzzy sets of input signals. For deriving these rules, expert knowledge and experience is generally used. \\3) \textit{Defuzzification}: this step reverts back the results from the fuzzy rule base in step (2) back to crisp mode  and activates the output action.

\begin{figure*}
\centering
\includegraphics[scale = 0.55]{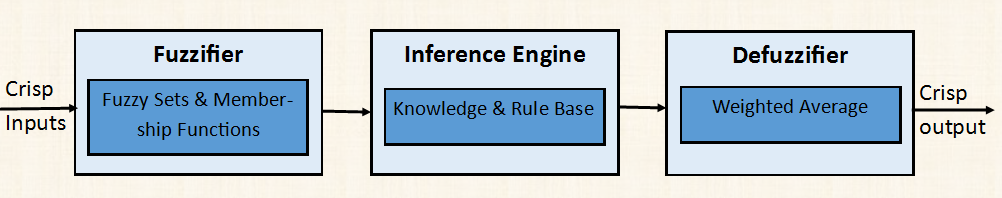}
\caption{FIS elements} 
\label{fig:fis}
\end{figure*}

FIS elements are shown in Fig. \ref{fig:fis}. The limitations of FIS arise from the rule base definition in step 2). FIS requires an expert knowledge to define the consequents of each rule (each combination of input signals and fuzzy sets). However, most of the consequents can not be easily deduced using previous knowledge/experience which can result in lower performance of the FIS. To overcome these challenges, FIS is applied in combination with RL, in particular Q-learning, to learn the consequents of each rule. This is described in detail in section \ref{sec:fql}.

\label{sec:algo}

\subsection{QL based controller}
Q-Learning is an off-policy RL algorithm that can learn the optimal Q-values for each state-action pairs \cite{suttonRL}. \emph{For the single agent case,} as long as all state-action pairs are visited and continued to be updated, Q-learning guarantee an optimal behavior regardless of the specific policy being followed throughout the learning phase. On the other hand, the multi agent case does not have a formal demonstration for the convergence to the optimal solution, due to the problem of conflict among the agents. 

The equation for updating the Q-values is given by:
\begin{equation}
\begin{split}
 Q(\boldsymbol{X}_t,A_t) = Q(\boldsymbol{X}_t,A_t) + \alpha (r_t  + \gamma \max_{\boldsymbol{A}}Q(\boldsymbol{X}_{t+1},A) - Q(\boldsymbol{X}_t,A_t))
\label{eq:q}
\end{split}
\end{equation}
where $\alpha$ is the learning rate, $\gamma$ is the discount factor, $A_t$ is the current action, $r_t$ is the immediate reward, $\boldsymbol{X}_t$ and $\boldsymbol{X}_{t+1}$ are the current and the next state respectively. The procedure of Q-learning algorithm is shown in Algorithm~\ref{alg:algorithmq}.
%\subsubsection{QL algorithm details}
In what follows, we describe the definition of states, reward and actions for QL based controller for solving our MDP.
\begin{enumerate}
\item \textit{States:}
According to the system model defined in Section~\ref{sec:network-model}, a state $\boldsymbol{X}_t= \{\boldsymbol{H}^t, \boldsymbol{B}^t, \boldsymbol\rho^t, \boldsymbol{L}^t\}$ is composed of the energy arrival $\boldsymbol{H}^t$, the battery level at the vSC $\boldsymbol{B}^t$, the traffic load at MBS $\boldsymbol\rho^t$ and the traffic load at vSC $\boldsymbol{L}^t$. 
%Hence at beginning of time slot, $t$, the state is given by $\boldsymbol{X}_t= \{\boldsymbol{H}^t, \boldsymbol{B}^t, \boldsymbol\rho^t, \boldsymbol{L}^t\}$, where $\boldsymbol{H}^t$ is the level of energy arrival at $t$, $\boldsymbol{B}^t$  is the battery level at slot $t$, $\boldsymbol\rho^t$ is the traffic load of MBS at slot $t$ and $\boldsymbol{L}^t$ is the traffic load at each vSCs with in their coverage. 
The values of each state variables 
%energy arrival ($\boldsymbol{H}^t$), battery  ($\boldsymbol{B}^t$), traffic load at MBS ($\boldsymbol\rho^t$) and traffic load at vSC ($\boldsymbol{L}^t$ ) 
are all normalized and quantized into $5$ levels. Hence, the QL based controller has $5\times5\times5\times5 = 625$ states. Since our optimization objective is system-wide where as the vSCs are taking actions in distributed fashion, the coordination among vSCs is achieved by including the MBS traffic load information in the states of the controller at each vSCs. In this way, vSCs action selection can be coordinated towards minimizing the MBS load which, in turn, is equivalent to minimizing the grid energy consumption.
% \emph{(MM: This is a multi-agent RL problem, I would add more information on this before, in the theoretic part (and also introducing it in the intro section), you can have some hints in my journal on LL to this respect.)}\\

\item \textit{Actions:} The set of possible actions are the possible operative mode of each vSC $\boldsymbol{A}^t$. Hence, the action set are combinations of the operative mode of each vSC which can be switched off, PHY-RF split mode and MAC-PHY split mode.

\item \textit{Reward:} The reward function determines the immediate reward the controller acquire as a result of taking a specific action. The optimization goal is to minimize the power drained from the grid while avoiding system outage as given by (\ref{eq:cost}). Hence, the reward function can be formulated as:
\begin{equation}
r_t = 1 -(\omega_1\cdot \mathrm{E_m}(\boldsymbol{A}^t,t)+ \omega_2\cdot {\mathrm{D}}(\boldsymbol{A}^t,t))
\label{eq:reward}
\end{equation}
where $\mathrm{E_m}(\boldsymbol{A}^t,t)$ and  $\mathrm{D}(\boldsymbol{A}^t,t)$ are respectively the normalized grid energy consumption and the traffic drop rate in the cluster, given the operative modes of the vSCs and the time step $t$. 
\end{enumerate}
\begin{algorithm}[H]
\caption{Q-Learning Algorithm }
\begin{algorithmic} 
\State Initialize $Q(X,A) \forall X\in \boldsymbol{X}, A \in \boldsymbol{A}$ arbitrarily 
\For {each episode}:
\State Initialize $\boldsymbol{X}_t$
\For {each step, $t$, of episode}:
\State Choose $A_t$ from $\boldsymbol{X}_t$ using policy derived from  Q
\State Take action $A_t$, get reward $r_t$ and next state $\boldsymbol{X}_{t+1}$
\State update Q-value using (\ref {eq:q})
\State $\boldsymbol{X}_t$ = $\boldsymbol{X}_{t+1}$
\EndFor
\EndFor
\end{algorithmic}
\label{alg:algorithmq}
\end{algorithm}

\subsection{FQL based controller}
\label{sec:fql}

The main advantage of RL algorithms is that they do not need a model of an environment, which makes it suitable to be applied for our study. However, QL can be inefficient for large state-action spaces and cannot be directly applied in problems involving continuous state-action spaces. In such cases, fine grained discretization of the state-action space helps, but at a cost of an exponential increase in the state space, which makes the learning process slow. In order to overcome these limitations, fuzzy functions approximation can be used with QL and achieve a more smooth action transition in response to a smooth change in states, without the need for fine grained discretization. This motivates us to design a fuzzy approximation based controller for our MDP. 

FQL allows to integrate the benefits of FIS in Q-learning: provide good approximations of the Q-function and enable the use of Q-learning in continuous state spaces \cite{glorennec1994fuzzy}. In FQL, let $\boldsymbol{X}$ be the crisp set of inputs defining the state of a learning agent. Crisp values are the exact values of the inputs without any form of preprocessing. The process of converting the crisp values to fuzzy values is known as fuzzification. Fuzzification is done according to the degree of membership determined from membership functions. Each fuzzy rule corresponds to a state and its firing strength defines the degree to which the agent is in that state. Unlike FIS, in FQL, rules do not have fixed consequents 
%(PD: You mean consequences?). 
The consequents of each rule are learned through exploration/exploitation algorithm. The resulting FIS will have competing actions for each rule and each rule with have the following form:

if $\boldsymbol{X}$ is $X_i$ then $A[i,1]$  with  $q[i,1]$\\
\hspace*{12ex}  or $A[i,j]$  with  $q[i,j]$\\
\hspace*{12ex}                      . \\
\hspace*{12ex}                  .\\
 \hspace*{12ex}                      .\\
   \hspace*{10ex}                  or    $A[i,k]$  with  $q[i,k]$,\\
where $A[i,k]$ is the $k^{th}$ possible action in rule $i$ and $q[i,k]$ its corresponding $q$ value. 

Each fuzzy rule is corresponding to a state. A state $X_i$ is defined as:  ($x_1$ is $X_{i,1}$ and $x_2$ is $X_{i,2}$ and .... $x_n$ is $X_{i,n}$), where $X_{i,j}$, $ i = 1,...,n$  and $ j=1,...,n$ are the fuzzy sets corresponding to the membership functions of each crisp inputs $x_i$. The exploration/exploitation algorithm chooses random actions (exploration) or actions with maximum $q$ values (exploitation). The procedure of FQL algorithm is shown in Algorithm~\ref{alg:algorithmfql}.

\begin{algorithm}[H]
\caption{Fuzzy Q-Learning Algorithm}
\begin{algorithmic} 
\State Initialize q-values $q(i,k)$ for each rule $i$ and number of possible actions $A_k$
\State Observe the crisp input state $\boldsymbol{X}_t$ 
\State Select action $A_i$ from the number of possible actions for each rule $i$ according to exploration/exploitation policy:\\
 \hspace{1cm}  $A_i$ = $argmax_j$ $q(i,j)$  with probability $1- \epsilon$,\\
 \hspace{1cm}  $A_i$ = $random$ ${A_j, j = 1, ... k}$ with probability $\epsilon$
\State Determine the global action $A(\boldsymbol{X}_t)$ for the state $\boldsymbol{X}_t$ using (\ref{eq:flc})
\State Estimate the corresponding Q value $Q(\boldsymbol{X}_t,A)$ using  (\ref{eq:qvalue})
\State Take action $A(\boldsymbol{X}_t)$ and observe the new state $\boldsymbol{X}_{t+1}$
\State Get the reward $r_t$ 
\State Estimate the value of the new state $v(\boldsymbol{X}_{t+1})$ using (\ref{eq:statevalue})
\State Calculate the error signal $\Delta Q$ using (\ref{eq:errorvalue})
\State Update q-values using (\ref{eq:qupdate})
\end{algorithmic}
\label{alg:algorithmfql}
\end{algorithm}

\begin{equation}
A(\boldsymbol{X}_t) = \dfrac{\sum_{i=1}^{i = N} w_i(\boldsymbol{X}_t)A_i}{\sum_{i=1}^{i=N} w_i(\boldsymbol{X}_t) } 
\label{eq:flc}
\end{equation}
where $w_i(\boldsymbol{X}_t)$ is the firing strength of rule $i$ which is determined by the membership functions of crisp input $\boldsymbol{X}_t$ using fuzzy \textit{and} operation and $A_i$ is the corresponding action/consequent of rule $i$ from the exploration/exploitation policy.

\begin{equation}
Q(\boldsymbol{X}_t,A) = \dfrac{\sum_{i=1}^{i = N} w_i(\boldsymbol{X}_t)A_iq(i, A_i)}{\sum_{i=1}^{i=N} w_i(\boldsymbol{X}_t) } 
\label{eq:qvalue}
\end{equation}
where $q(i,A_i)$ is the q-value associated with rule $i$ and its selected action $A_i$.

\begin{equation}
v(\boldsymbol{X}_{t+1}) = \dfrac{\sum_{i=1}^{i = N} w_i(\boldsymbol{X}_{t+1})A_iq(i, A_{max})}{\sum_{i=1}^{i=N} w_i(\boldsymbol{X}_{t+1}) } 
\label{eq:statevalue}
\end{equation}
where $w_i(\boldsymbol{X}_{t+1})$ is the firing strength of rule $i$ evaluated from the new state $\boldsymbol{X}_{t+1}$  and $q(i, A_{max})$ is the maximum q-value for rule $i$.
\begin{equation}
\Delta Q = r_t + \gamma v(\boldsymbol{X}_{t+1}) - Q(\boldsymbol{X}_t,A)  
\label{eq:errorvalue}
\end{equation}
where $\gamma$ is the discount factor.

\begin{equation}
\Delta q(i, A_i)  = \alpha \Delta Q  \dfrac{w_i(\boldsymbol{X}_t)}{\sum_{i=1}^{i=N} w_i(\boldsymbol{X}_{t}) } 
\label{eq:qupdate}
\end{equation}
where $\alpha$ is the learning rate.

%\subsubsection{FQL Algorithm Details}
In our scenario, we define the membership functions for the traffic load, energy arrival and battery as well as actions and reward functions, as follows.
\begin{enumerate}
\item \textit{Membership Functions and Fuzzy Rules:}
The crisp input state is $\boldsymbol{X}_t$, defined in Section~\ref{sec:network-model}. 
Trapezoidal and triangular membership functions are used for the traffic load at MBS, traffic load at vSCs, energy arrival and battery level of vSCs. In particular, $5$ fuzzy sets are defined with linguistic terms "Very Low", "Low", "Medium", "High" and "Very High" as shown in Fig. \ref{fig:membership}. Hence, the fuzzification step involves mapping the input $\boldsymbol{X}_t$ into $5$ fuzzy sets for traffic load at MBS, traffic load at each vSCs, energy arrivals and battery level of each vSCs. Hence, there are $625$ rules corresponding to every combination of fuzzy sets in the FQL. Similarly as for QL-based controller design, MBS traffic load information is included in the fuzzy rules definition of each vSCs controller to achieve coordination among vSCs towards a common optimization goal of minimizing grid energy consumption while avoiding system outages. 
\begin{figure}
\centering
\includegraphics[scale = 0.5]{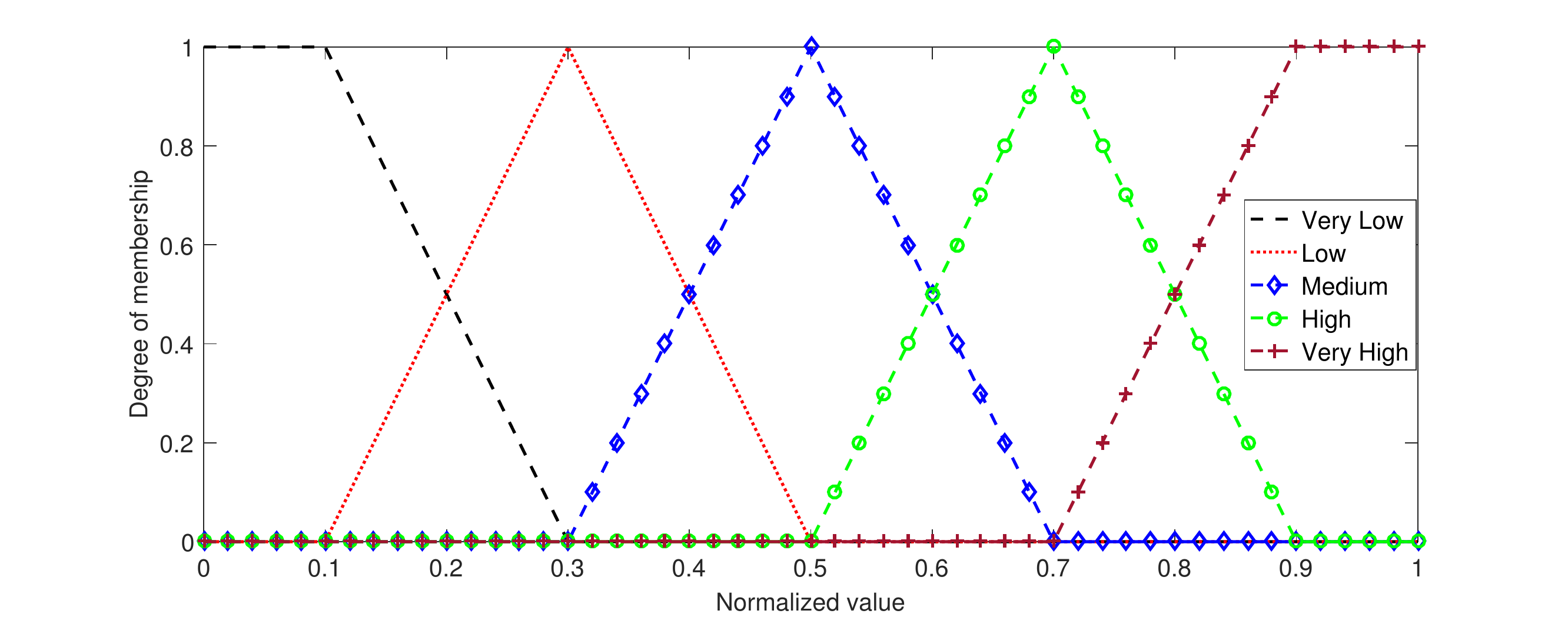}
\caption{Membership functions of MBS traffic load, vSC traffic load, energy harvesting and battery level} 
\label{fig:membership}
\end{figure}
\item \textit{Actions:}
The set of possible actions are the possible operative mode of each vSC. 
%Hence, the action set are combinations of the operative mode of each vSC which can be switch-off, PHY-RF split mode and MAC-PHY split mode. 
An action for each rule is determined by using the exploration/exploitation policy as shown in step $2$ of Algorithm~\ref{alg:algorithmfql}. After computing the firing strength of each rule using membership functions, the global action is computed as a weighted sum of each action and the corresponding firing strength using (\ref{eq:flc}). This defuzzification method is commonly applied in zero order Sugeno fuzzy systems \cite{sugeno} and is known to be computationally efficient. In our controllers, since the set of actions are limited ($3$ operative modes of each vSC), the crisp output obtained by the defuzzification method using (\ref{eq:flc}) is converted to a nearest integer, which corresponds to an operative mode of a vSC.
\item \textit{Reward:}
The reward function is the same as defined for QL control in (\ref{eq:reward}).
\end{enumerate}
%determines the immediate reward the controller acquire as a result of taking a specific action. The optimization goal is to minimize the power drained from the grid while avoiding system outage as given by (\ref{eq:cost}). Hence, the reward function can be formulated as:
%\begin{equation}
%r_t = 1 -(\omega_1\cdot \mathrm{E_m}(\boldsymbol{A}^t,t)+ \omega_2\cdot {\mathrm{D}}(\boldsymbol{A}^t,t))
%\label{eq:reward}
%\end{equation}
%where $\mathrm{E_m}(\boldsymbol{A}^t,t)$ and  $\mathrm{D}(\boldsymbol{A}^t,t)$ are respectively the normalized grid energy consumption and the traffic drop rate in the cluster, given the operative modes of the vSCs and the time step $t$. 
\subsection{Control without coordination}
\label{ssec:u-control}

This section presents the control methods where each vSCs take actions independently without any system wide information. In this case, as opposed to the QL and FQL methods described above, the vSCs did not have the load level of the MBS and it is not considered in their decision making process. As a result the state/rules of un-coordinated methods are given by:

\begin{equation}
 \boldsymbol{X}_t= \{\boldsymbol{H}^t, \boldsymbol{B}^t, \boldsymbol{L}^t\}
\label{eq:states-un}
\end{equation}

Therefore, the un-coordinated methods have $5\times5\times5 = 125$ rules/states for FQL and QL methods respectively. We call these methods as Un-Coordinated FQL (U-FQL) and Un-Coordinated QL (U-QL). The actions and reward definitions of U-FQL and U-QL methods are the same to the actions and rewards defined above for both FQL and QL controls.

\section {Numerical Results}

\label{sec:results}

\subsection{Simulation Scenario}
According to the traffic model defined in Section~\ref{sec-eh-traffic-profiles}, user activities are categorized based on~\cite{auer2010} as heavy users with an activity of $900$ MB/hr and ordinary users with an activity of $112.5$ MB/hr. 
The solar energy traces are generated using the SolarStat tool~\cite{Miozzo2014} for the city of Los Angeles. For the PV modules, we have considered the commercial Panasonic N235B. These panels have single cell efficiencies as high as 21.1\%, which ranks them amongst the most efficient solar modules at the time of writing, delivering about 186 $\mathrm{W/m^2}$. The solar panel size and battery capacity are dimensioned based on the criteria that the vSC can be fully recharged on a typical winter day. The simulation parameters and reference power consumption values are given in Table~\ref{tab:table1}. The BB static power consumption figures are composed of $P_\mathrm{CPU}$, $P_\mathrm{OFDM}$ and $P_\mathrm{filter}$ and the load dependent components are $P_\mathrm{FD}$ and $P_\mathrm{FEC}$.

%The reference vSC power consumption values for our scenario are $P_\mathrm{RF}=2.6$ W and $P_\mathrm{PA}=71.4$ W. For $P_{BB}$, we consider $200$ GOPS, $160$ GOPS and $80$ GOPS for $P_{\mathrm{CPU}}$, $P_{\mathrm{filter}}$ and  $P_{\mathrm{OFDM}}$ respectively. Moreover, the reference load dependent power consumption values are  $30$ GOPS, $10$ GOPS and $20$ GOPS for linear component of $P_{\mathrm{FD}}$, non-linear component of $P_{\mathrm{FD}}$ and $P_{\mathrm{FEC}}$ respectively. 
%As for the MBS, we consider $P_\mathrm{RF}=9.18$ W and $P_\mathrm{PA}=1100$ W. The baseband power consumption is 630 GOPS and 215 GOPS for the static ($P_\mathrm{CPU}+P_\mathrm{OFDM}+P_\mathrm{filter}$) and load dependent components ($P_\mathrm{FD}+P_\mathrm{FEC}$), respectively. The power consumption overhead ($P_{\mathrm{overhead}}$) is of 0 and 10\% of the total power of the rest of the base station for the case of vSC and MBS, respectively.\emph{(MM: I would create a table with all the parameters, in order to give to the reader a unique reference for them)}. %\emph{(MM: I though we were using the optimal value based on DP)}.

\begin{table}[ht]
  \centering
  \caption{Simulation Parameters.}
  \begin{tabular}{l c}
   Parameter & Value\\
   \hline \hline
   Transmission power of macro cell (dBm) & 43\\
   Transmission power of vSC (dBm) & 38\\
   Bandwidth (MHz) & 5\\
   MIMO Transmission Mode  & 2x2\\
   UEs per vSC & 90\\
   Heavy users ratio & 0.5\\
   Solar panel size ($\mathrm{m}^2$) & 4.48 \\
   Battery capacity (kWh) & 2\\
   $B_{th}$ & $20\%$\\
   $P_{RF_{vSC}}$ & $2.6$ W\\
   $P_{PA_{vSC}}$ & $71.4$ W\\
   $P_{RF_{MBS}}$ & $9.18$ W\\
   $P_{PA_{MBS}}$ & $1100$ W\\
   GOPS to W conversion factor & 8\\
   $P_{BB_{{static}_{vSC}}}$ & $440$ GOPS\\
   $P_{BB_{{load-dependent}_{vSC}}}$ & $60$ GOPS\\
   $P_{BB_{{static}_{MBS}}}$ & $630$ GOPS\\
   $P_{BB_{{load-dependent}_{MBS}}}$ & $215$ GOPS\\
   $P_{overhead_{vSC}}$ & $0.0\%$\\
   $P_{overhead_{MBS}}$ & $10.0\%$\\
   
  \hline
  \end{tabular}
\label{tab:table1}
\end{table}

\subsection{Off-line Training}
\label{sec:offline-training}
In this section we analyze the behavior of the system when the training is performed off-line. In particular, we considered one year as an episode with time granularity of one hour, since it allows to achieve a correct dimensioning of the solar power system for cellular base stations, as shown in~\cite{DaSilva2018}. Hence, every hour the agents choose actions corresponding to one of the possible operative modes of the vSCs with the goal of minimizing grid energy consumption, while avoiding system outage.
%The environment remains stable per year basis, the episode, in order to be able to evaluate progress in learning. 

\subsubsection{Training Analysis}

The training phase requires calibrating the parameters of the algorithm that have the strongest impact appropriately. These parameters are the learning rate ($\alpha$), the exploration parameter ($\epsilon$) and discount factor ($\gamma$). Moreover, we also adopt a discount process on these parameters in order to guide the exploration toward the stability. In particular, we are applying an exploration discount factor of $0.5$ at the beginning of each epoch until the agent reaches minimum level of exploration, which is equivalent to $3\%$.
%\emph{(MM: how they are discounted?)}. 

%[PD: FIG6-7 AND FIG8-9 CAN BE MERGED IN a) b) ON THE SAME LINE. PLEASE CHANGE ALSO THE SENTENCES INTRODUCING THE FIGURE ACCORDINGLY.]

\begin{figure}[h]
\centering
\captionsetup{position=auto}
\captionsetup[subfloat]{captionskip=-1pt}
\vspace*{-0.0cm}
 \subfloat[\label{cr-nsc3-residential-fql}]{\includegraphics[width=2.8in]{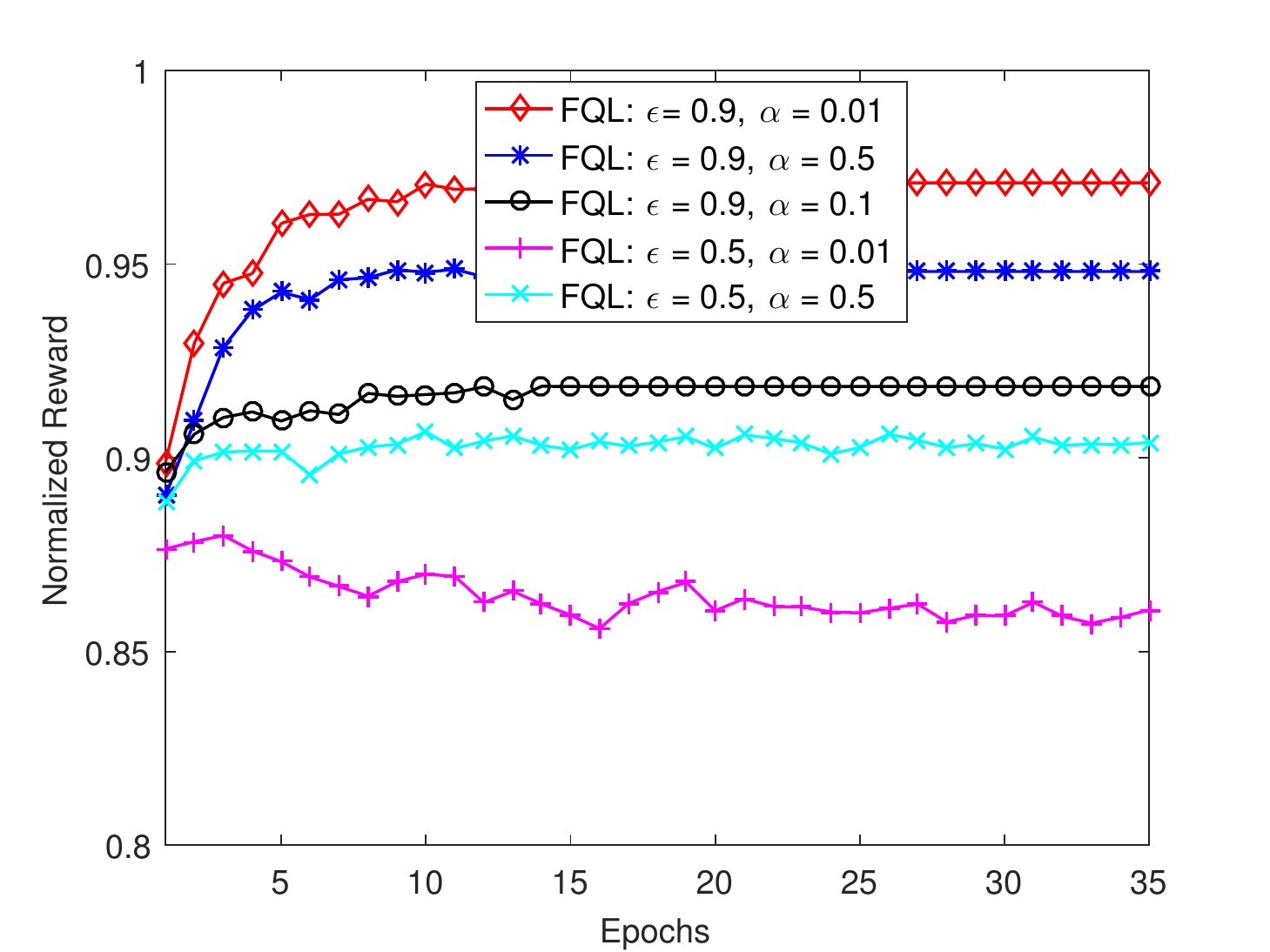}}\hspace{-0em}\quad
 \subfloat[\label{cr-nsc3-residential-ql}]{\includegraphics[width=2.8in]{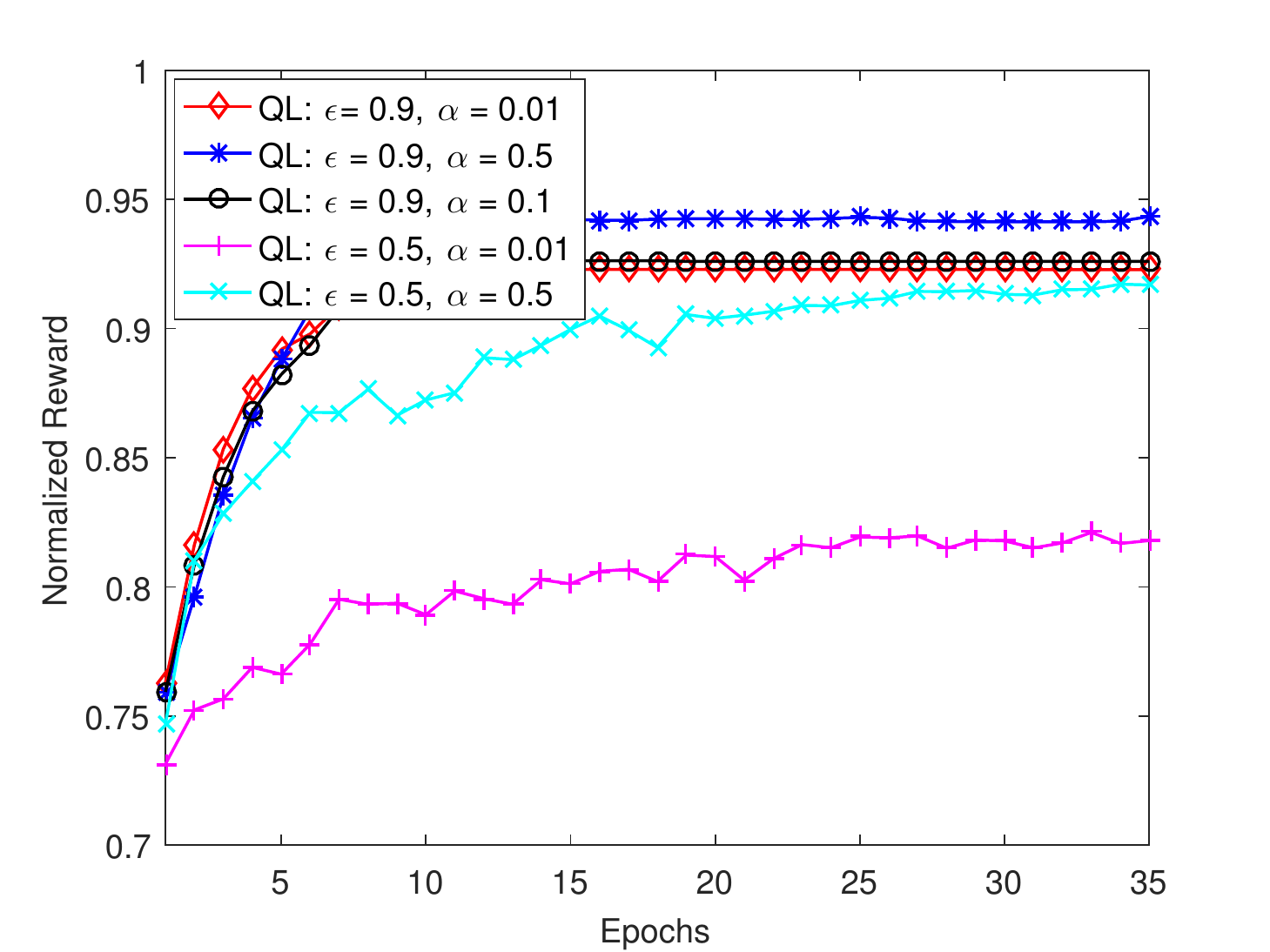}}\hspace{-0em}\quad
 \vspace*{-0.3cm}
\caption{Cumulative reward in residential area corresponding to different training parameters: (a)~FQL (b)~QL }
\label{fig:cr-nsc3-residential}
\end{figure}

The cumulative reward of FQL and QL methods for a system composed of $3$ vSCs and a MBS with in a residential area traffic profile are shown in Fig. \ref{fig:cr-nsc3-residential}. The cumulative reward is normalized with respect to a cumulative reward bound which is determined off-line using DP \cite{fss-vtc}. As it can be seen from Fig. \ref{cr-nsc3-residential-fql}, FQL based control can achieve a cumulative reward very close to the optimal bound (more than $97\%$). In addition, the choice of training parameters affects the convergence of the FQL control. The cumulative reward is shown to be sensitive to the exploration and learning rate parameter choices as it can reach the $85\%$ level in the case of $\alpha=0.01$ and $\epsilon=0.5$. The cumulative reward of QL based control in residential area is shown in Fig. \ref{cr-nsc3-residential-ql}. In the best case, the cumulative reward obtained by QL ($94\%$) is close to the optimal bound but lower with respect to FQL. 
%\emph{[MM: It is not clear to me whether with have a "discontinuos" behavior with alpha, e.g., 0.01 is working better wrt 0.5, which is behaving better of 0.1, any idea?]}
\comment{
\begin{figure}
\centering
\includegraphics[scale = 0.5]{cr_nsc3_residential_combined}
\caption{Cumulative reward of FQL in residential area corresponding to different training parameters} 
\label{fig:cr-nsc3-residential}
\end{figure}
\begin{figure}
\centering
\includegraphics[scale = 0.5]{cr_nsc3_office_combined.eps}
\caption{Cumulative reward of FQL in residential area corresponding to different training parameters} 
\label{fig:cr-nsc3-office}
\end{figure}
}

\begin{figure}[h]
\centering
\captionsetup{position=auto}
\captionsetup[subfloat]{captionskip=-1pt}
\vspace*{-0.0cm}
 \subfloat[\label{cr-nsc3-office-fql}]{\includegraphics[width=2.8in]{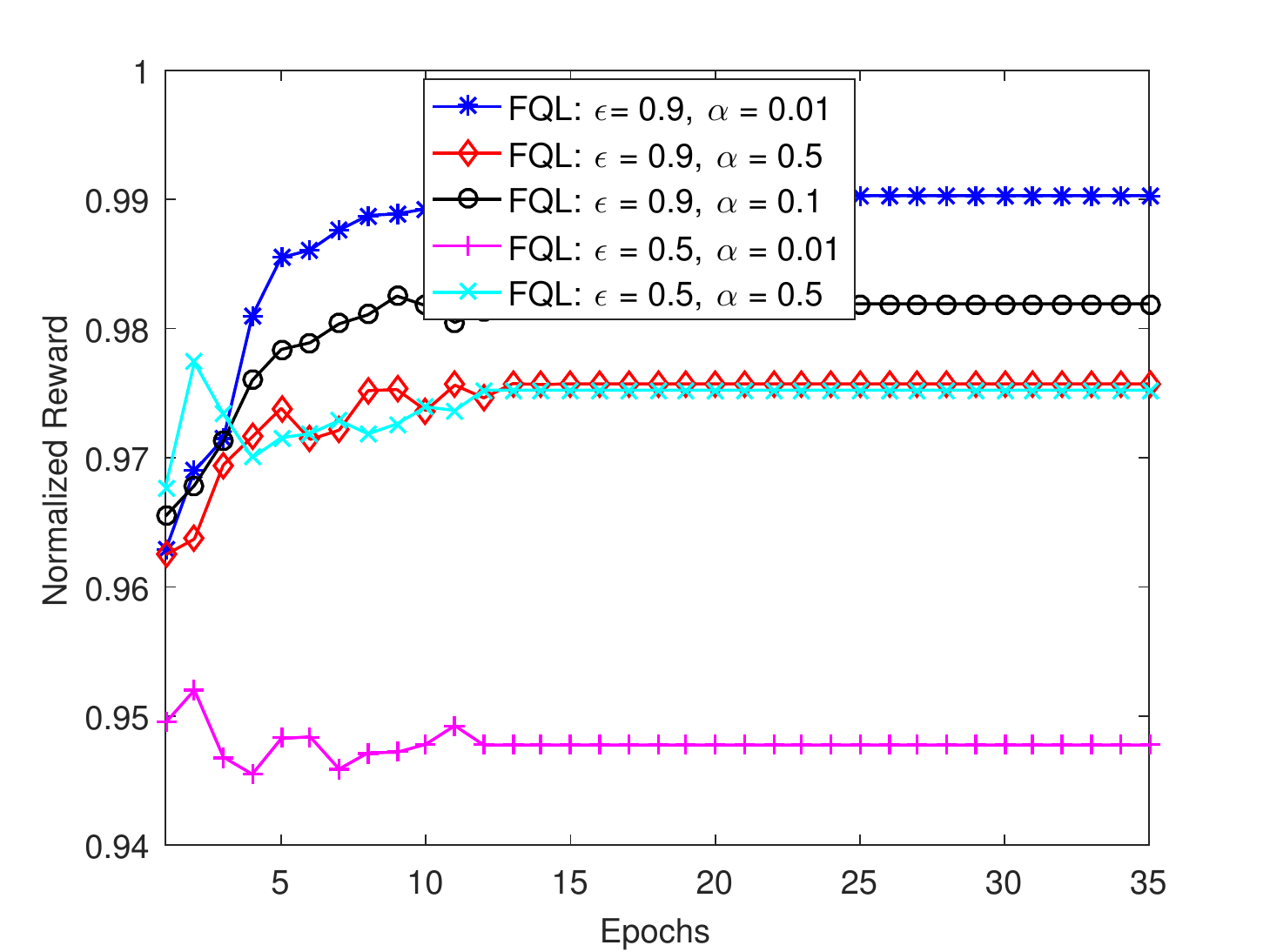}}\hspace{-0em}\quad
\subfloat[\label{cr-nsc3-officel-ql}]{\includegraphics[width=2.8in]{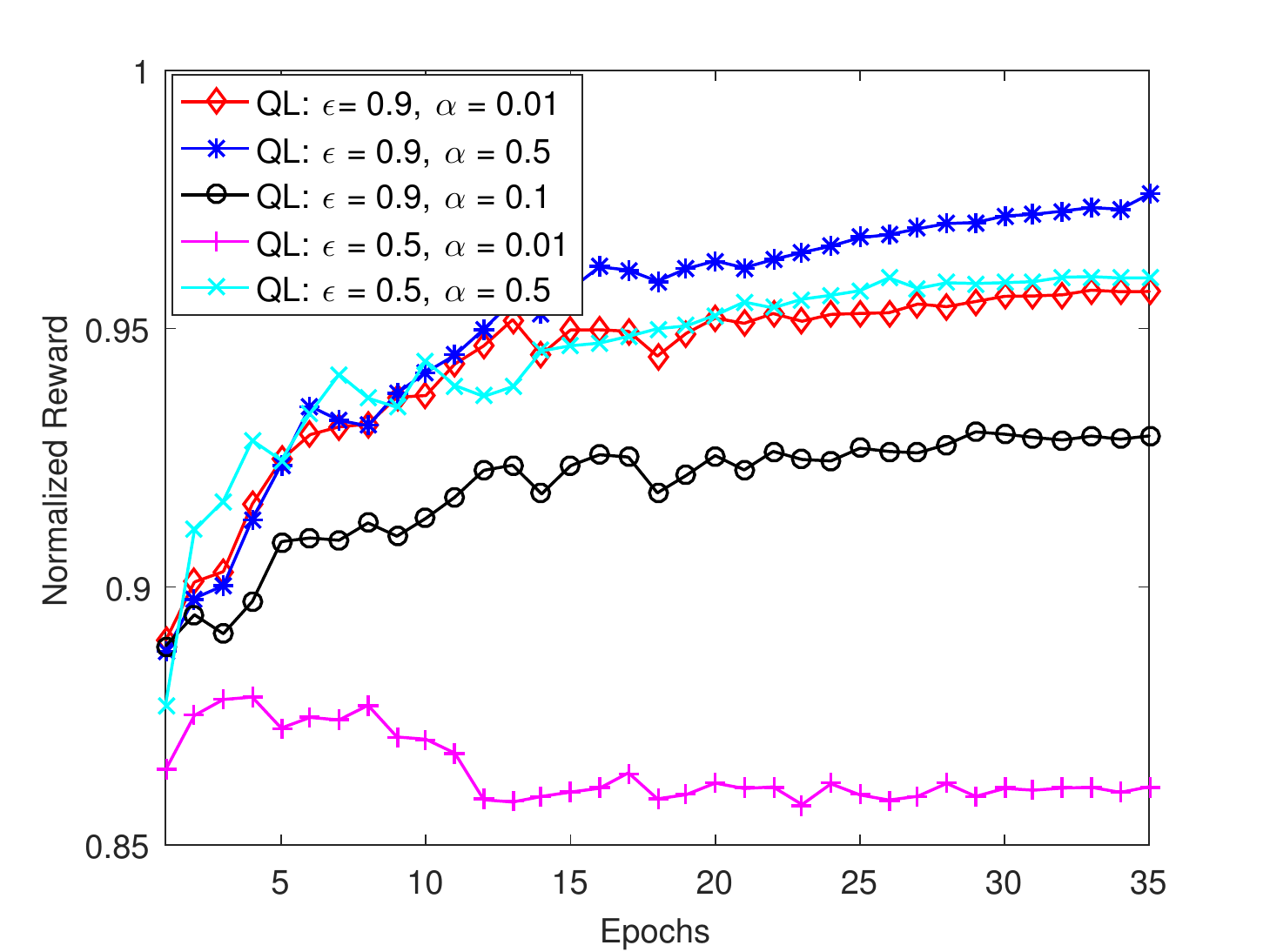}}\hspace{-0em}\quad
 \vspace*{-0.3cm}
\caption{Cumulative reward in office area corresponding to different training parameters: (a)~FQL (b)~QL }
\label{fig:cr-nsc3-office}
\end{figure}

\comment{
\begin{figure}
\centering
\includegraphics[scale = 0.5]{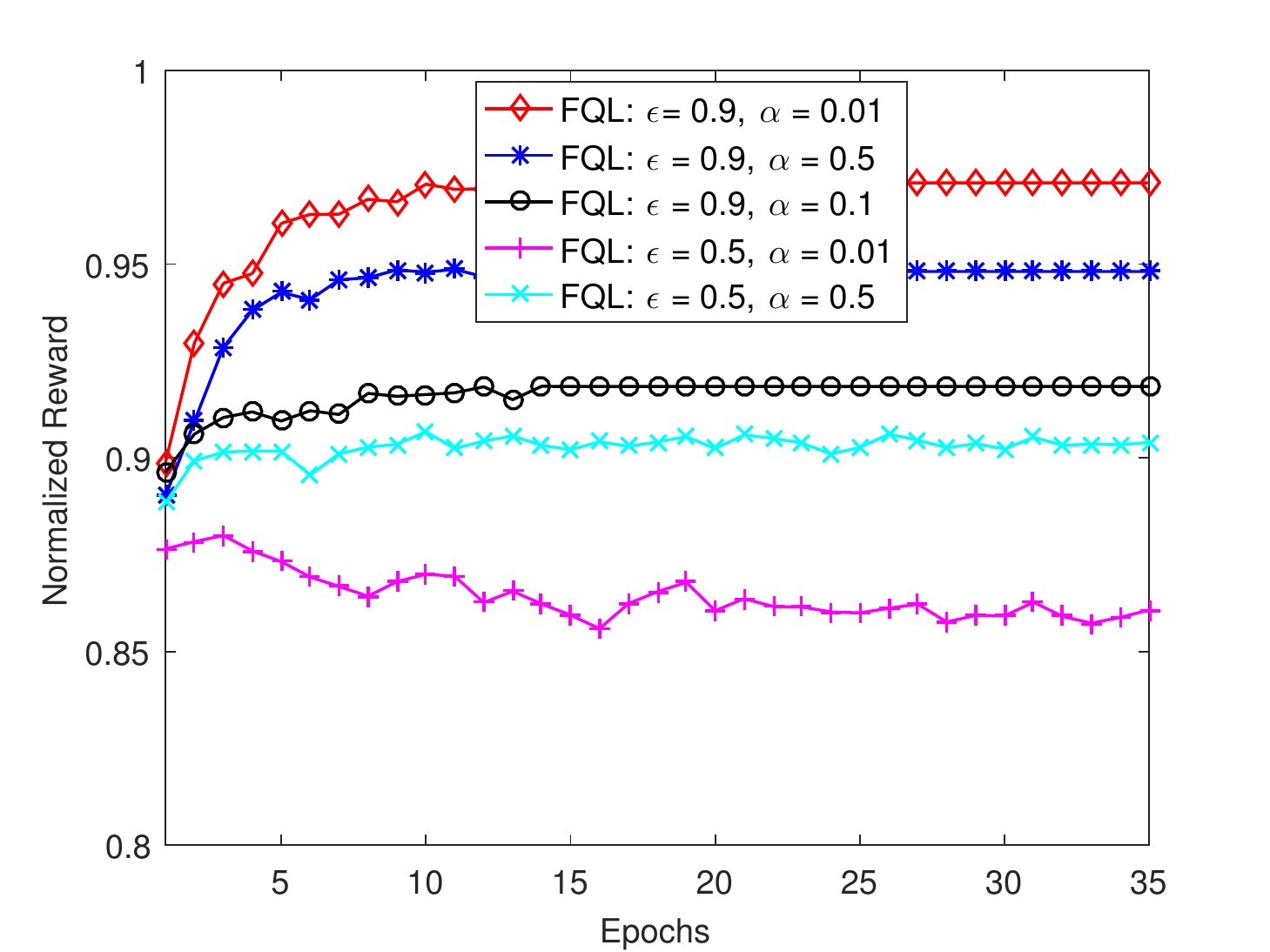}
\caption{Cumulative reward of FQL in residential area corresponding to different training parameters} 
\label{fig:yearlytraining-nsc3_fql}
\end{figure}
\begin{figure}
\centering
\includegraphics[scale = 0.5]{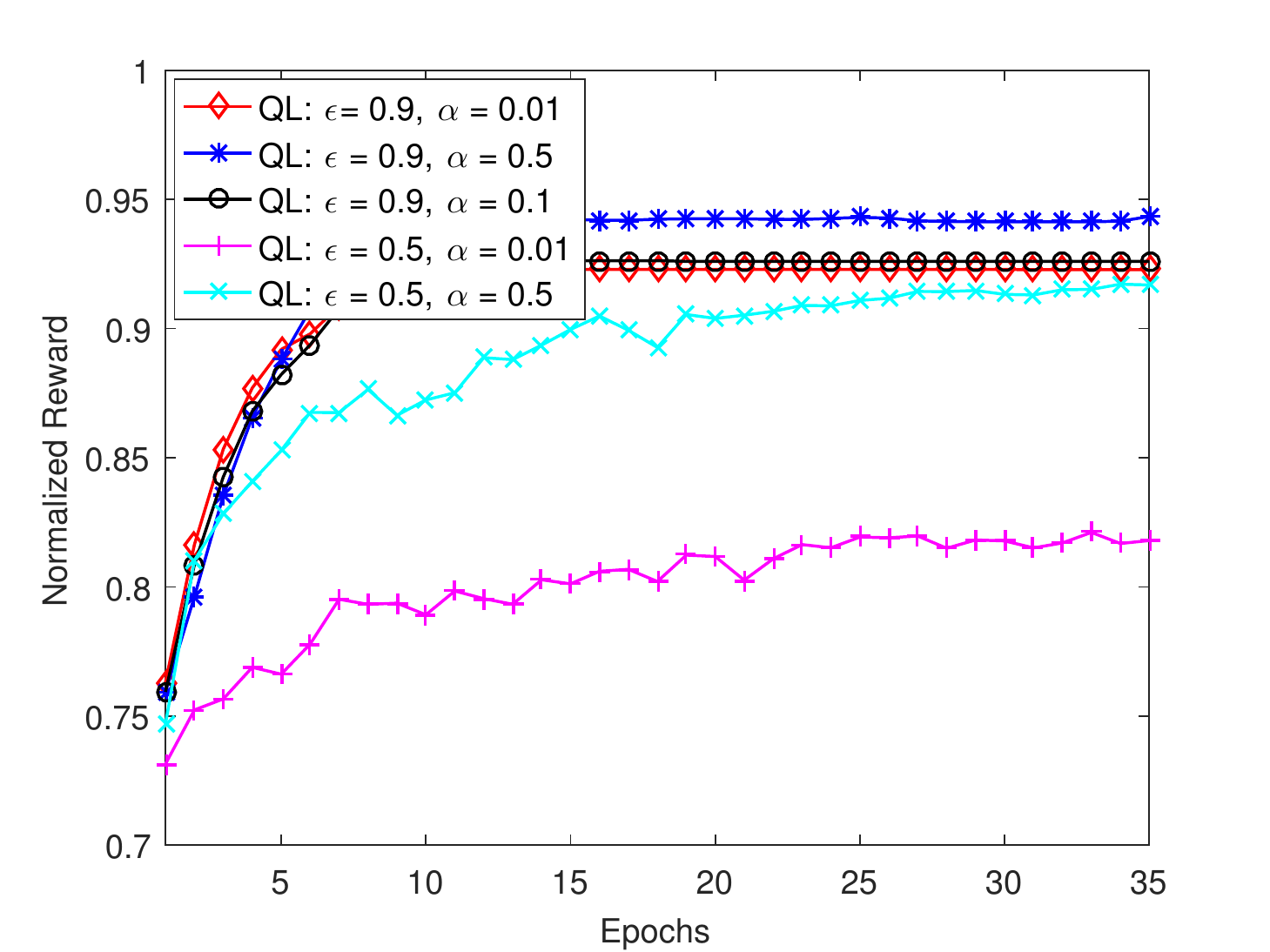}
\caption{Cumulative reward of QL in residential area corresponding to different training parameters} 
\label{fig:yearlytraining-nsc3_ql}
\end{figure}
\begin{figure}
\centering
\includegraphics[scale = 0.5]{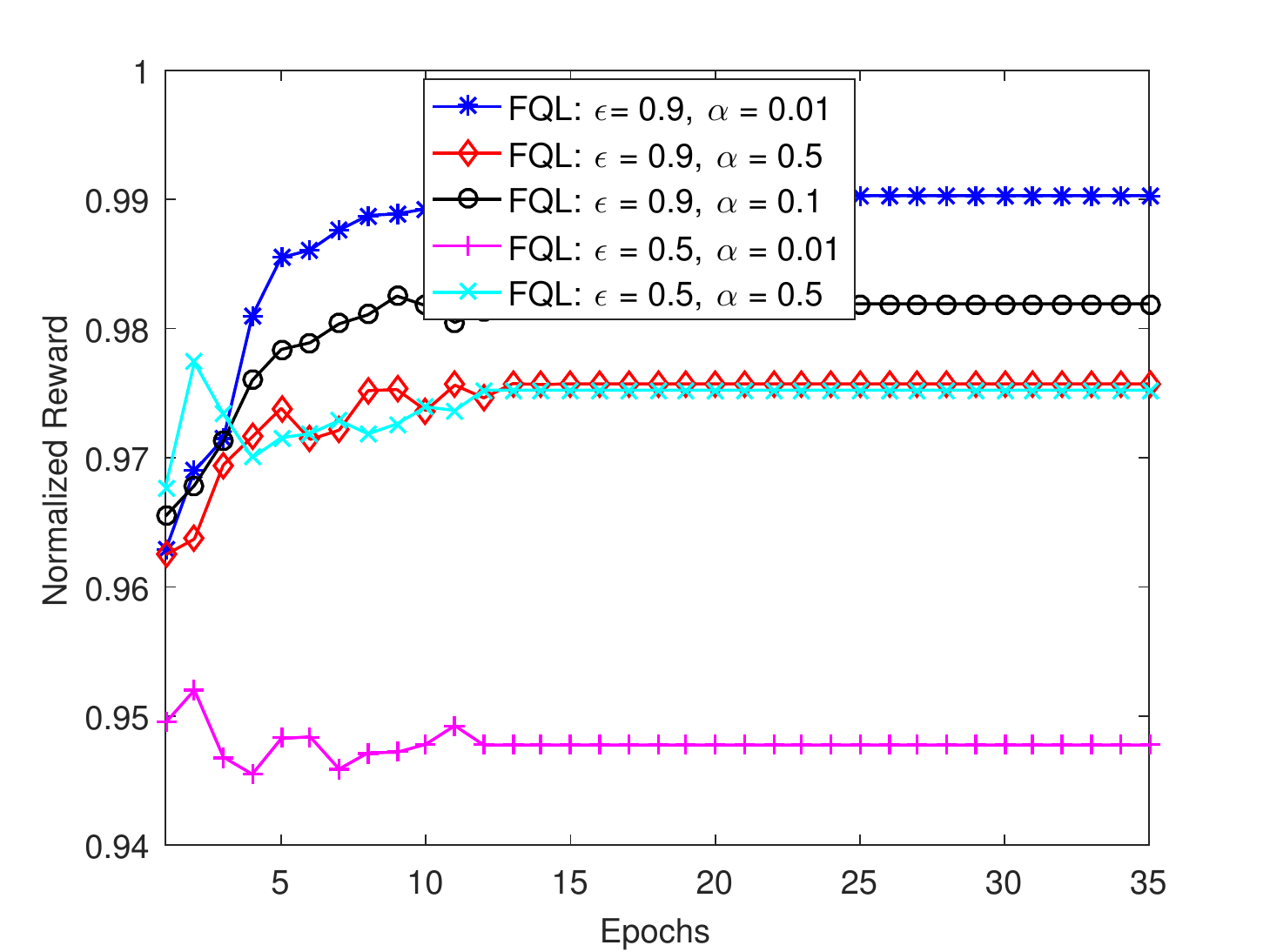}
\caption{Cumulative reward of FQL in an office area corresponding to different training parameters} 
\label{fig:yearlytraining-office-nsc3_fql}
\end{figure}
\begin{figure}
\centering
\includegraphics[scale = 0.5]{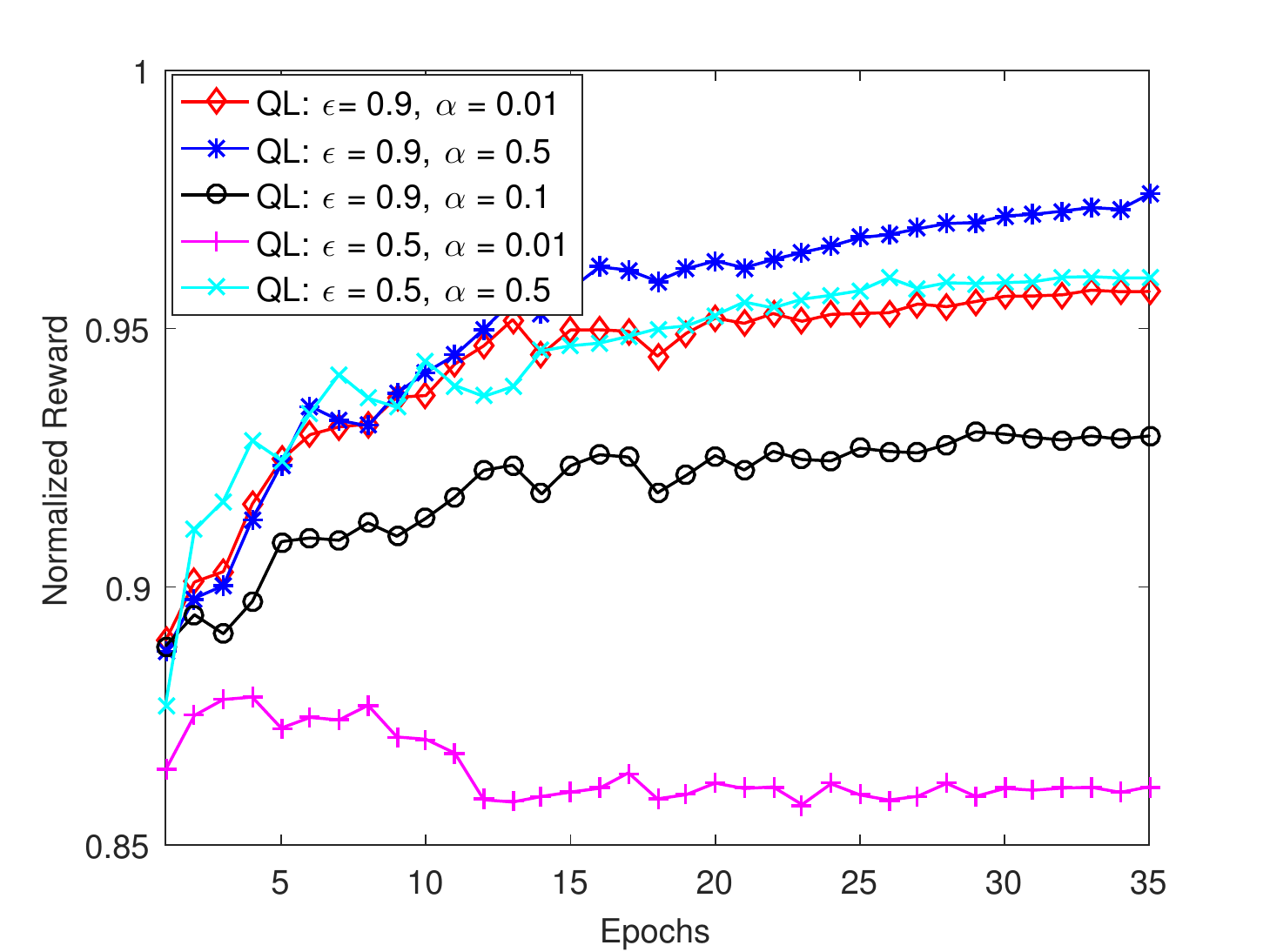}
\caption{Cumulative reward of QL in an office area corresponding to different training parameters} 
\label{fig:yearlytraining-office-nsc3_ql}
\end{figure}
}
The normalized cumulative reward for a system of $3$ vSCs deployed in an office area is shown in Fig. \ref{fig:cr-nsc3-office}. These results show that at the best case, FQL and QL controls in an office area are able to gain a cumulative reward of about $99\%$  and $97\%$ with respect to the optimal bound, respectively. The results in Fig. \ref{fig:cr-nsc3-residential} and Fig. \ref{fig:cr-nsc3-office} also show that, FQL based control is able to accumulate rewards faster than QL. In residential scenario, the FQL method is able to get around $95\%$ of the reward in less than $5$ epochs where as QL requires about $15$ epochs to reach the same level of cumulative reward. For an office scenario, FQL achieves a cumulative reward of about $97\%$ in less than $5$ epochs, whereas QL requires about $20$ epochs to reach $96\%$ level. Moreover, higher initial exploration rate is important for both FQL and QL, since it enables to explore more actions randomly during the initial phase of the training. Thus, the agent in the vSC has already discovered a higher number of rules-actions/state-actions pairs for FQL and QL, respectively, which,  in turn, help to avoid entering local optima.

\begin{figure}[h]
\centering
\captionsetup{position=auto}
\captionsetup[subfloat]{captionskip=-3pt}
\vspace*{-0.0cm}
\subfloat[\label{cr-bar-residential}]{\includegraphics[width=2.5in]{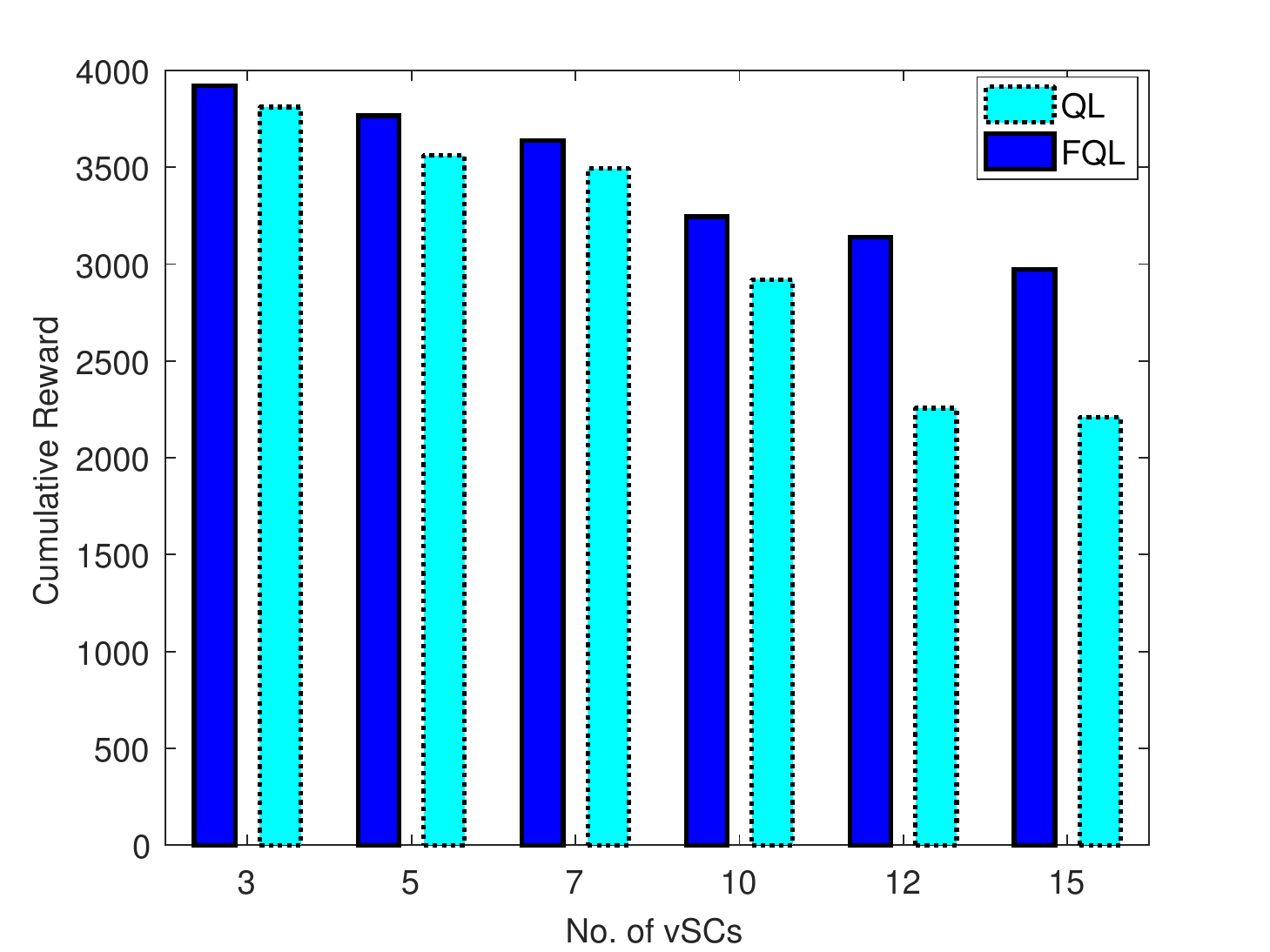}}\hspace{-0em}\quad
\subfloat[\label{cr-bar-office}]{\includegraphics[width=2.5in]{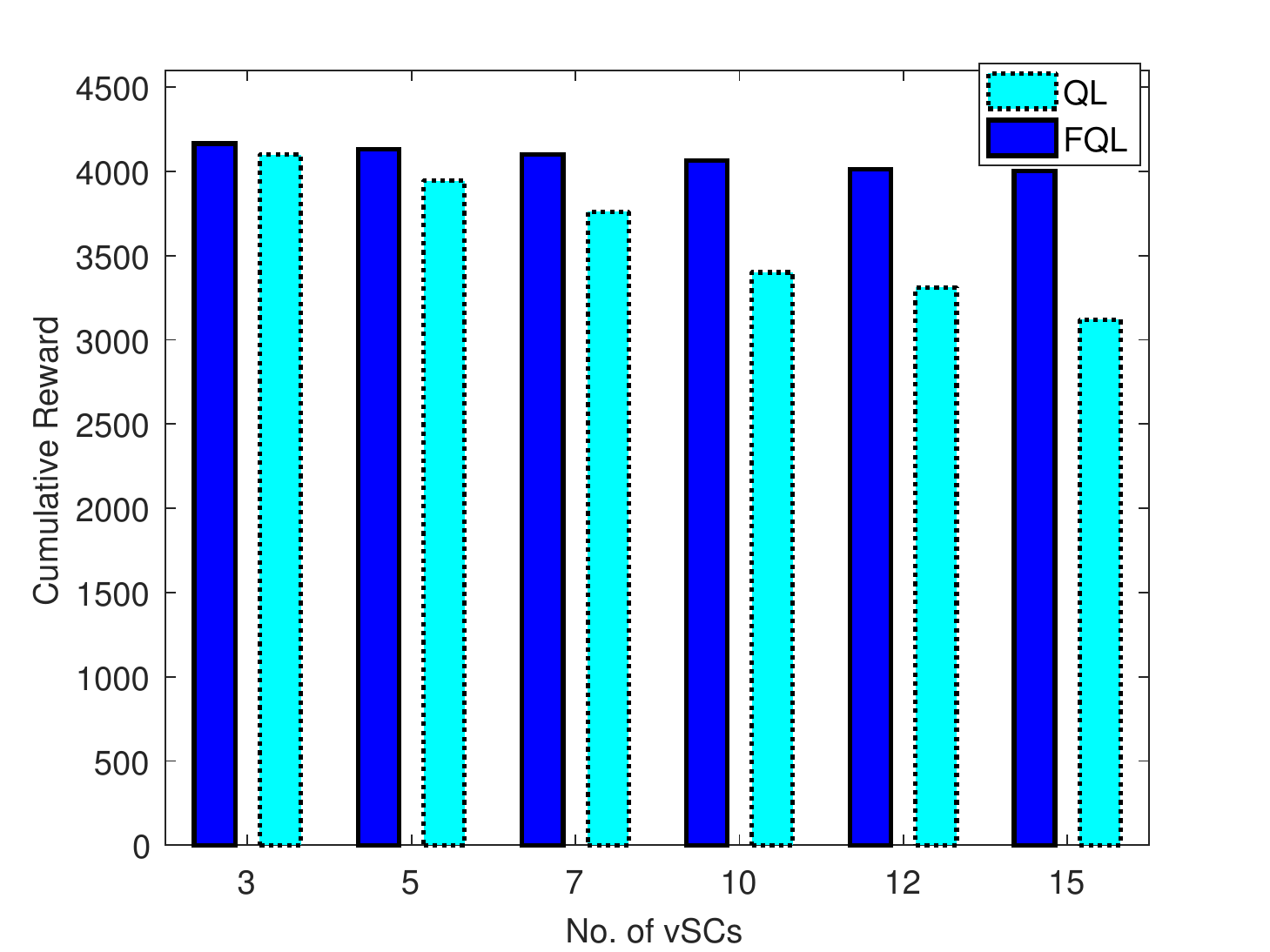}}\hspace{-0em}\quad
 \vspace*{-0.3cm}
\caption{Maximum cumulative reward obtained by FQL and QL for $3$, $5$, $7$, $10$, $12$ and $15$ vSCs (a)~Residential (b)~Office }
\label{fig:cr-bar}
\end{figure}

The same analysis has been also applied for scenarios with $5$, $7$, $10$, $12$ and $15$ vSCs. The maximum cumulative reward obtained by both QL and FQL based controllers in residential and office area are shown in Fig. \ref{fig:cr-bar}. The results show that FQL is able to accumulate higher reward compared to QL, $35\%$ more with $15$ vSCs. It is to be noted that, the maximum cumulative reward is decreasing as the number of vSCs increases. This is due to the higher load in the system injected by the vSCs which generates higher system drop rate and, in turn, reduces the immediate reward. Moreover, as the number of vSCs increases, conflicts among the actions of the agents can emerge which can impact the immediate reward obtainable.
%[PD: MAY THIS BEHAVIOR BE AN EFFECT OF THE CONFLICTING ACTIONS BY THE AGENTS TOO?].
 In addition, the cumulative reward is higher in an office area. In fact, the peak of traffic in the residential profile occurs during the early night (11 pm), as shown in Fig. \ref{fig:traffic-eh}, when the energy income is low, thus forcing the agents to switch-off or choose actions with more computational offloading to MBS. 
%As a result, the reward obtained in each slot in residential profile can be lower than the reward obtained in office traffic profile. 
Finally, the maximum cumulative reward gap between FQL and QL increases with the number of vSCs, as can be seen in  Fig. \ref{cr-bar-residential} and Fig. \ref{cr-bar-office}. This highlights the better suitability of FQL control especially in a network of higher number of vSCs.
%\emph{(MM: Why they are decreasing with a different slope?)}. 
\comment{
\begin{figure}
\centering
\includegraphics[scale = 0.5]{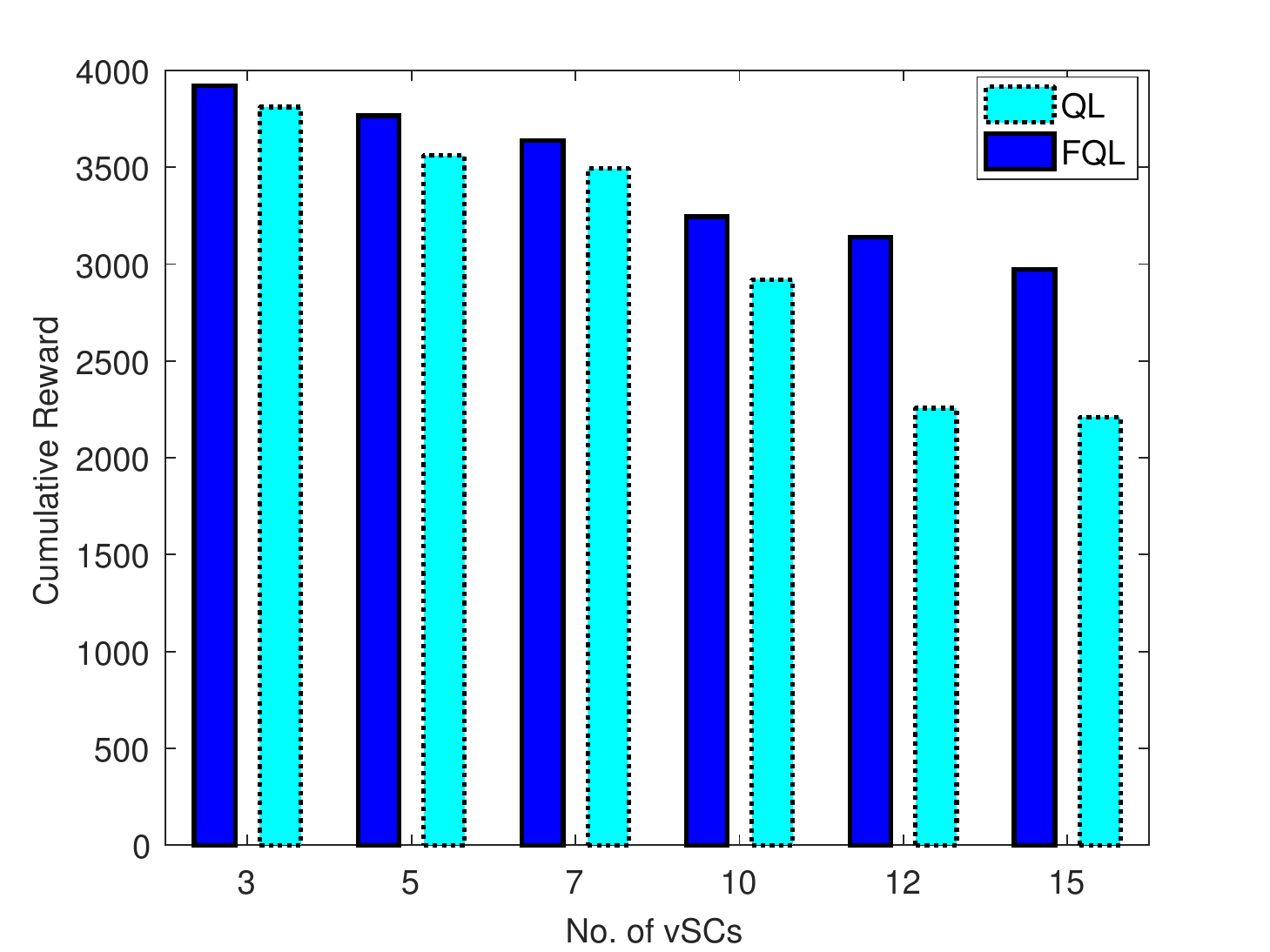}
\caption{Maximum cumulative reward obtained by FQL and QL in residential area for $3$, $5$, $7$, $10$, $12$ and $15$ vSCs} 
\label{fig:cr-residential}
\end{figure}
\begin{figure}
\centering
\includegraphics[scale = 0.5]{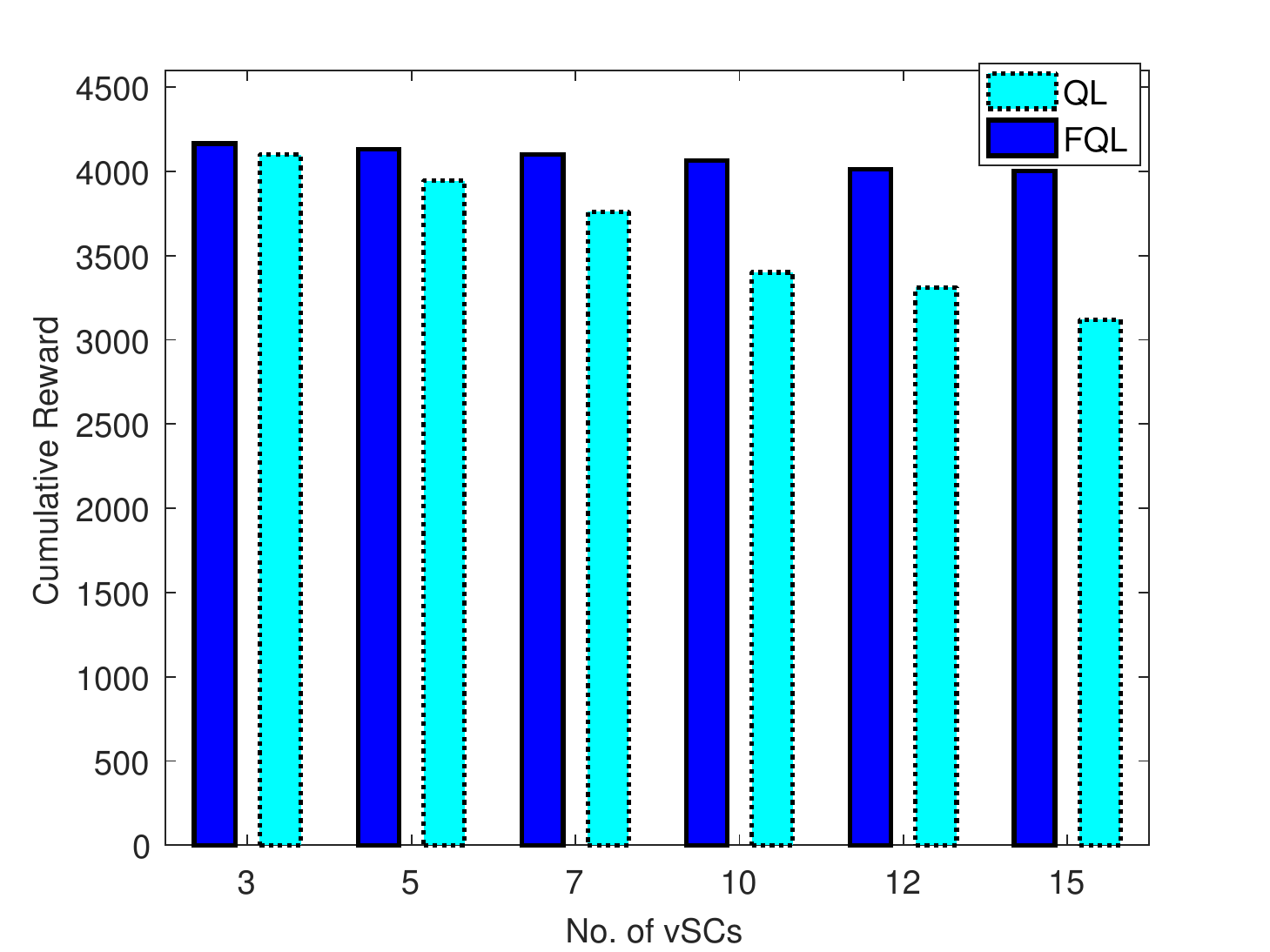}
\caption{Maximum cumulative reward obtained by FQL and QL in office area for $3$, $5$, $7$, $10$, $12$ and $15$ vSCs} 
\label{fig:cr-office}
\end{figure}
}

\subsubsection{Policy Characteristics}
\begin{figure}[h]
\centering
\captionsetup{position=top}
\captionsetup[subfloat]{captionskip=-3pt}
\vspace*{-0.0cm}
 \hspace*{-0.5cm}\subfloat[]{\includegraphics[width=2.35in]{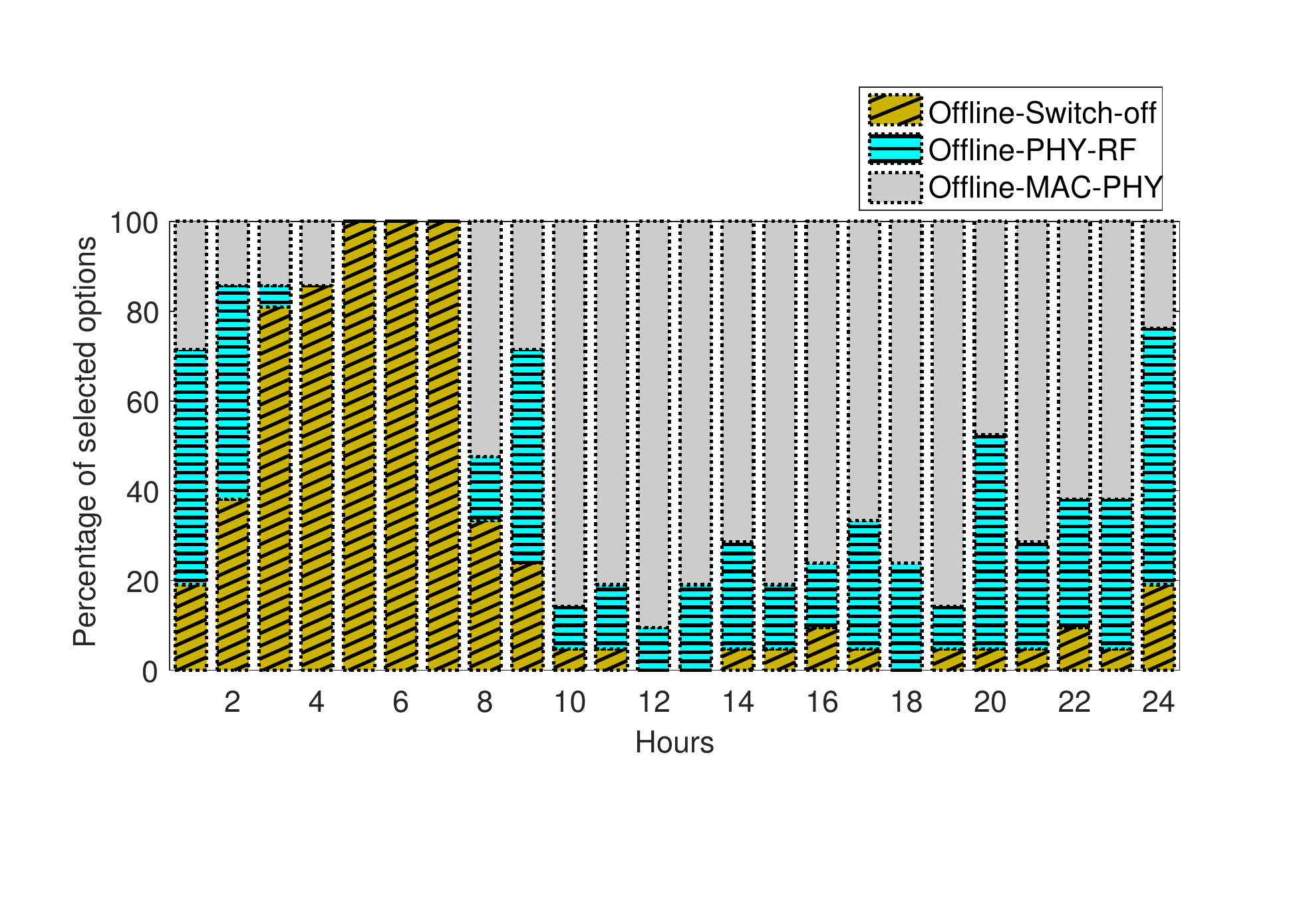}}\hspace{-2.2em}\quad
\subfloat[]{\includegraphics[width=2.35in]{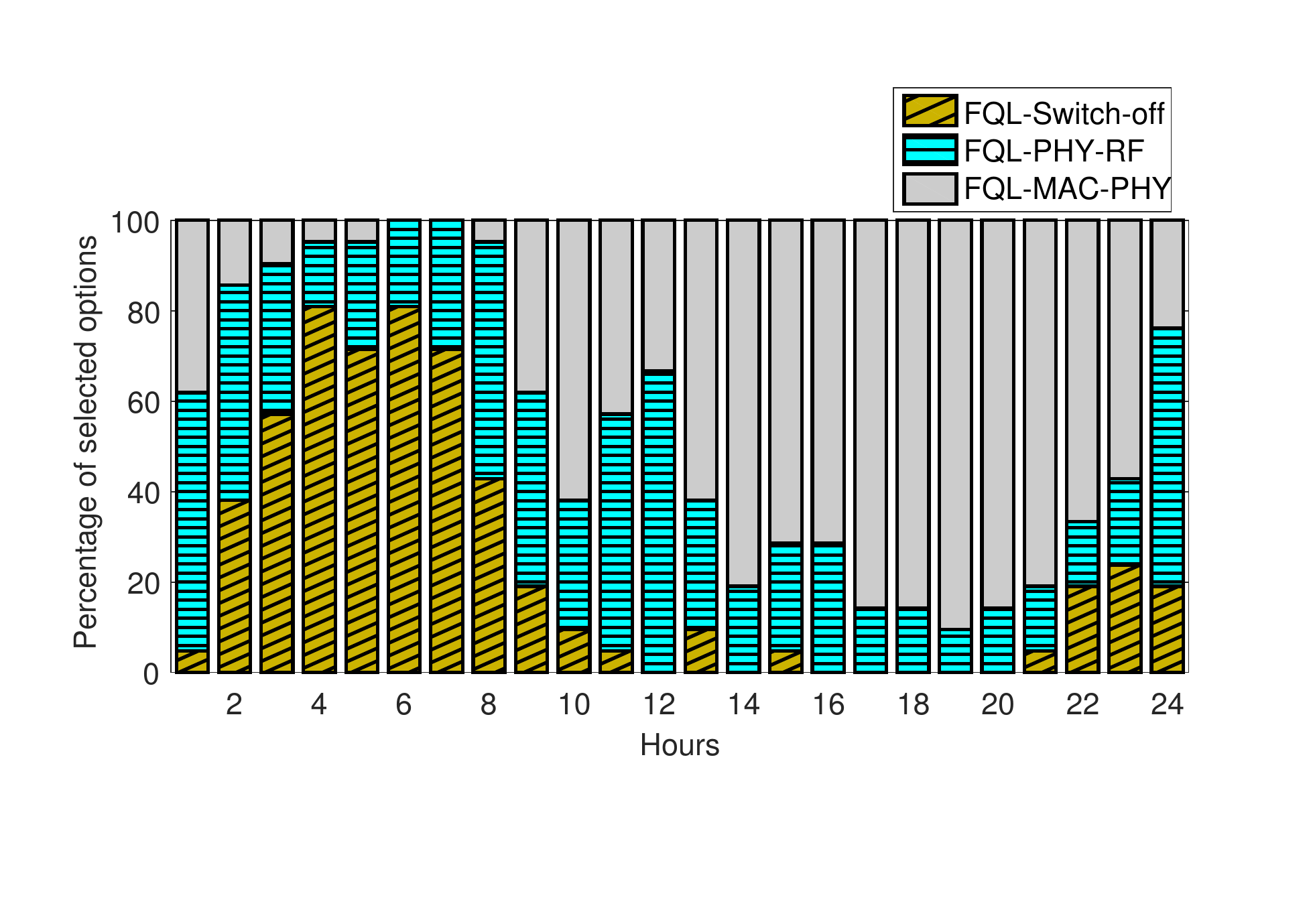}}\hspace{-2.2em}\quad
\subfloat[]{\includegraphics[width=2.35in]{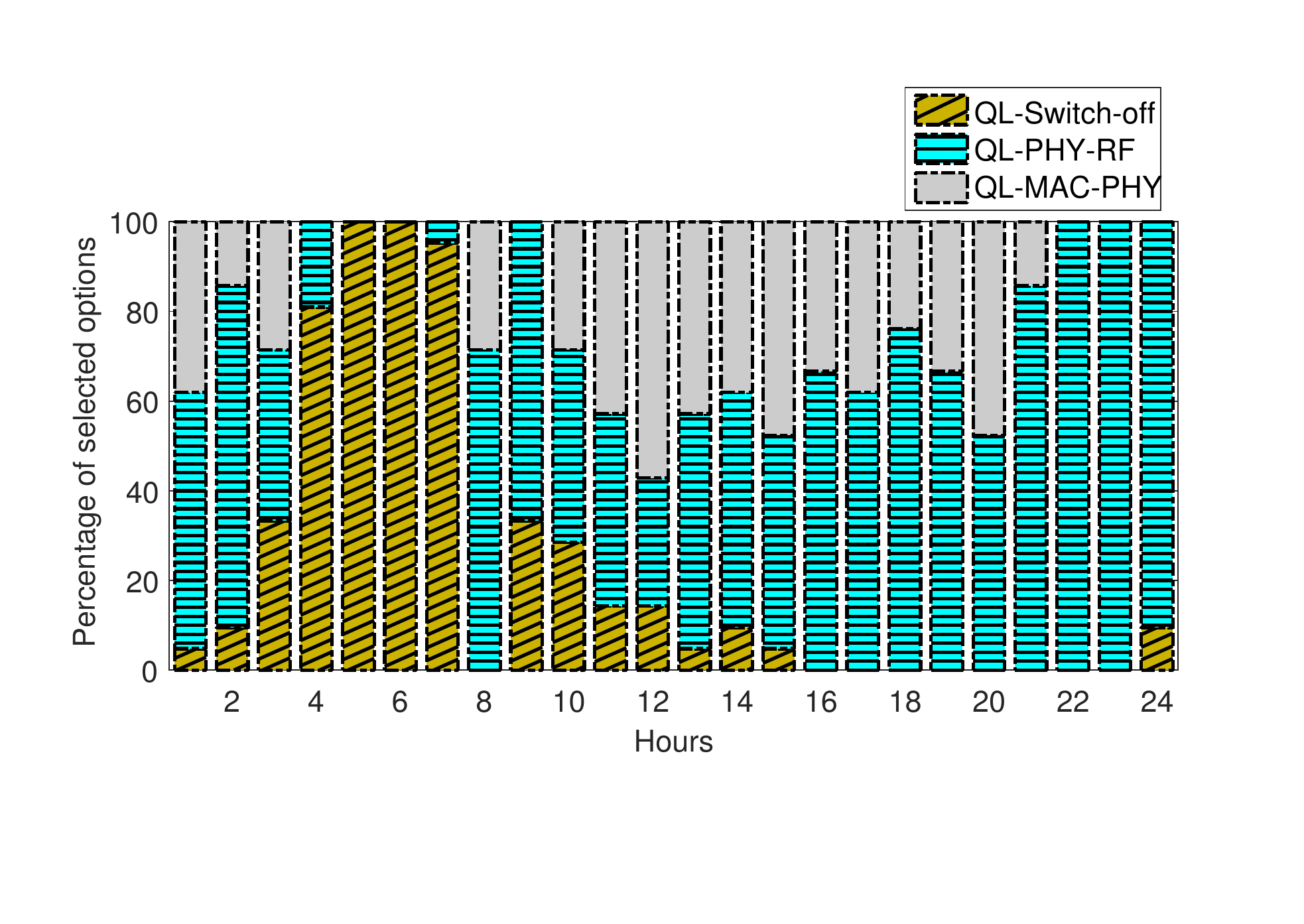}}\hspace{-2.0em}
 \vspace*{-0.6cm}
\caption{Average residential area winter day policy characteristics of: (a)~Off-line (b)~FQL (c)~QL }
\label{fig:winter-policy-residential}
\end{figure}

\begin{figure}[h]
\centering
\captionsetup{position=top}
\captionsetup[subfloat]{captionskip=-3pt}
\vspace*{-0.0cm}
 \hspace*{-0.5cm}\subfloat[]{\includegraphics[width=2.35in]{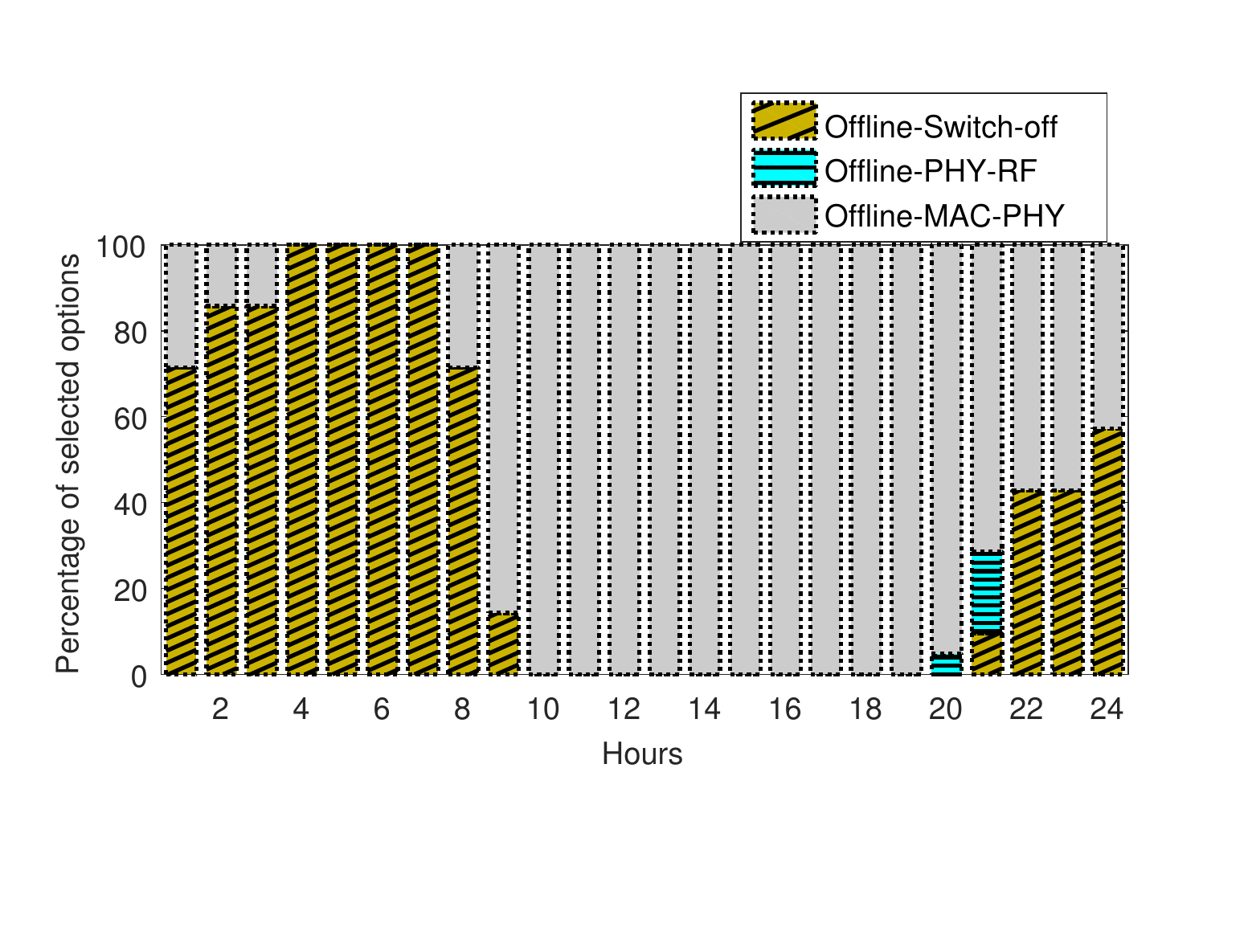}}\hspace{-2.2em}\quad
\subfloat[]{\includegraphics[width=2.35in]{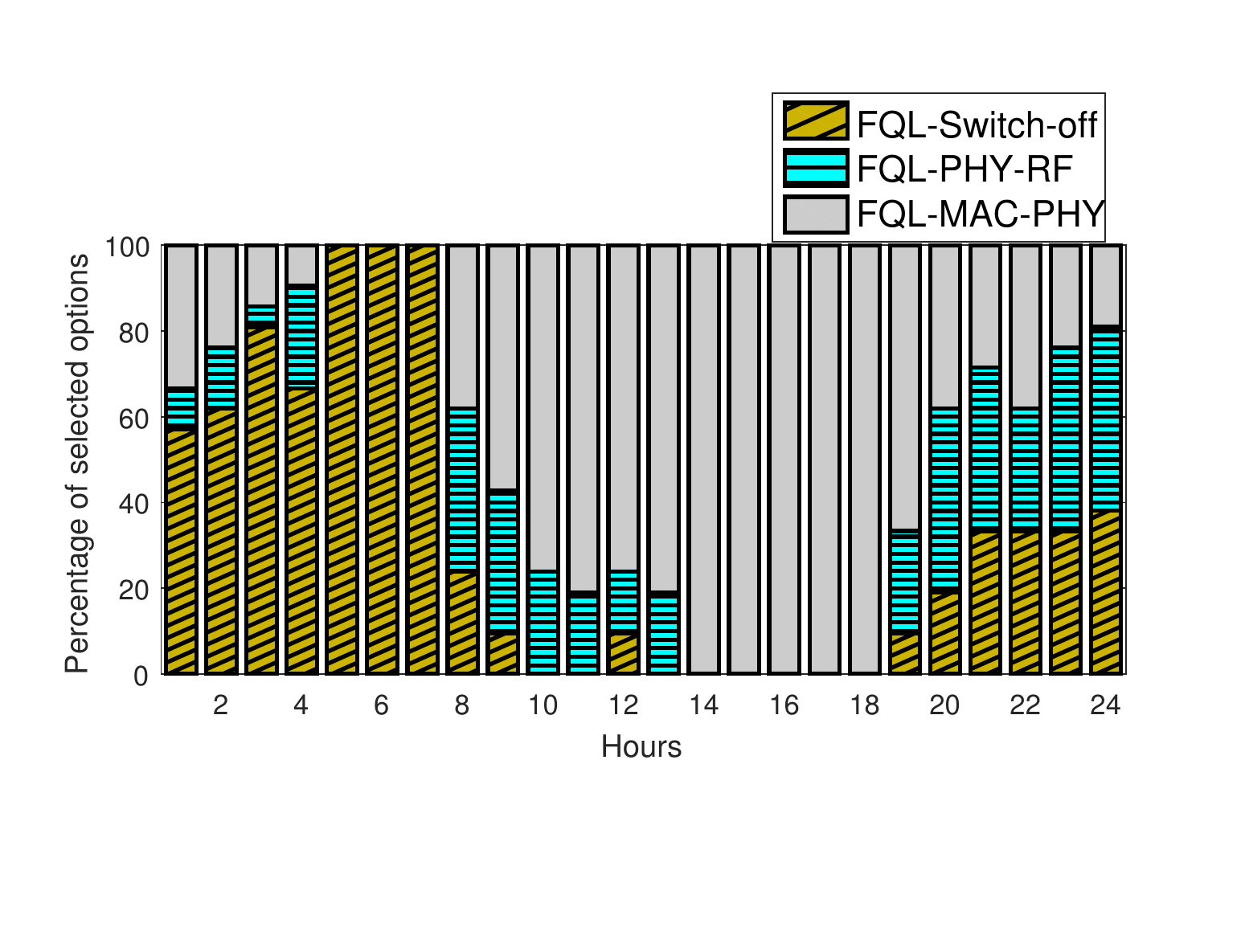}}\hspace{-2.2em}\quad
\subfloat[]{\includegraphics[width=2.35in]{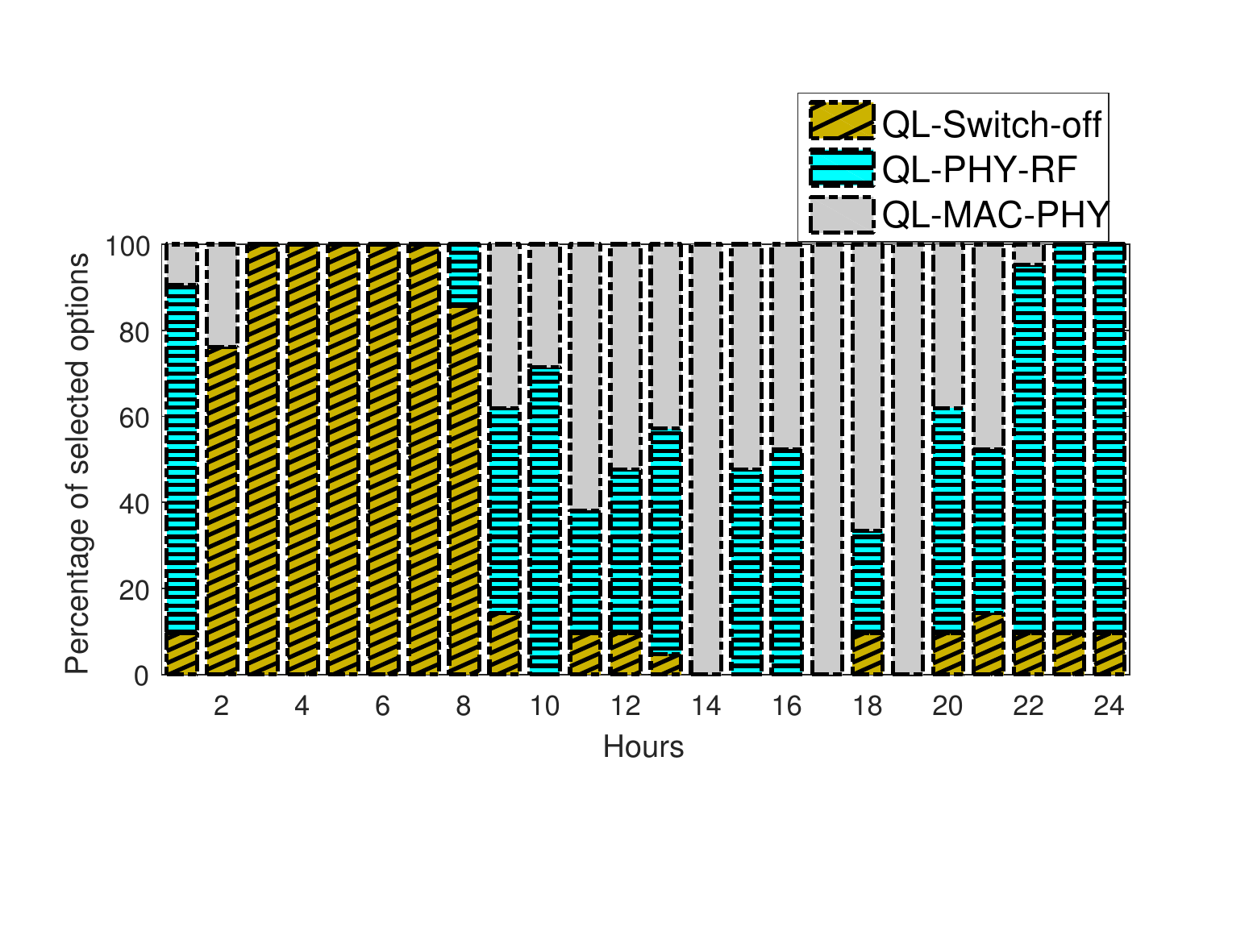}}\hspace{-2.0em}
 \vspace*{-0.6cm}
\caption{Average office area winter day policy characteristics of: (a)~Off-line (b)~FQL (c)~QL }
\label{fig:winter-policy-office}
\end{figure}

\begin{figure}[h]
\centering
\captionsetup{position=top}
\captionsetup[subfloat]{captionskip=-3pt}
 \hspace*{-0.5cm}\subfloat[]{\includegraphics[width=2.35in]{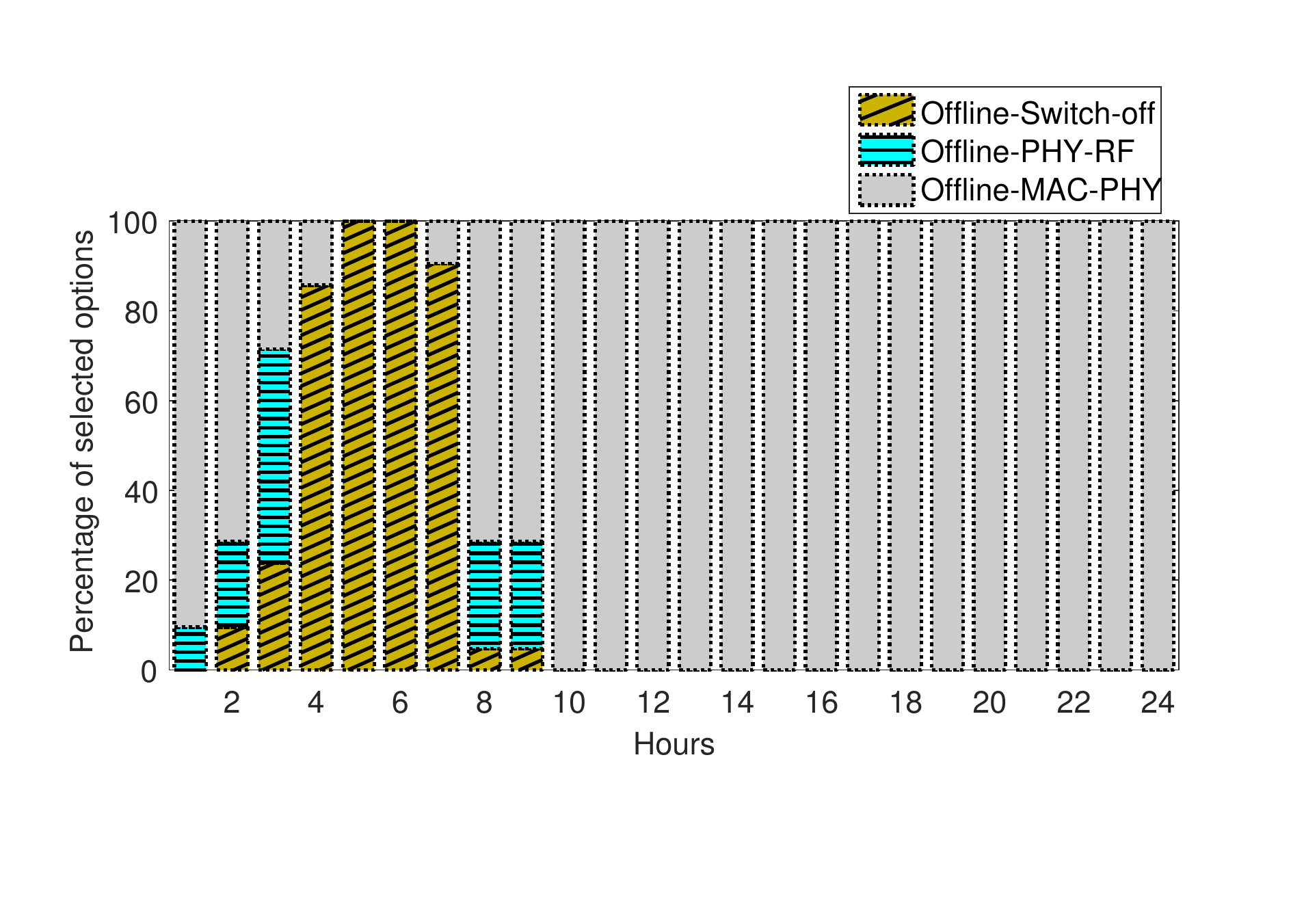}}\hspace{-2.2em}\quad
\subfloat[]{\includegraphics[width=2.35in]{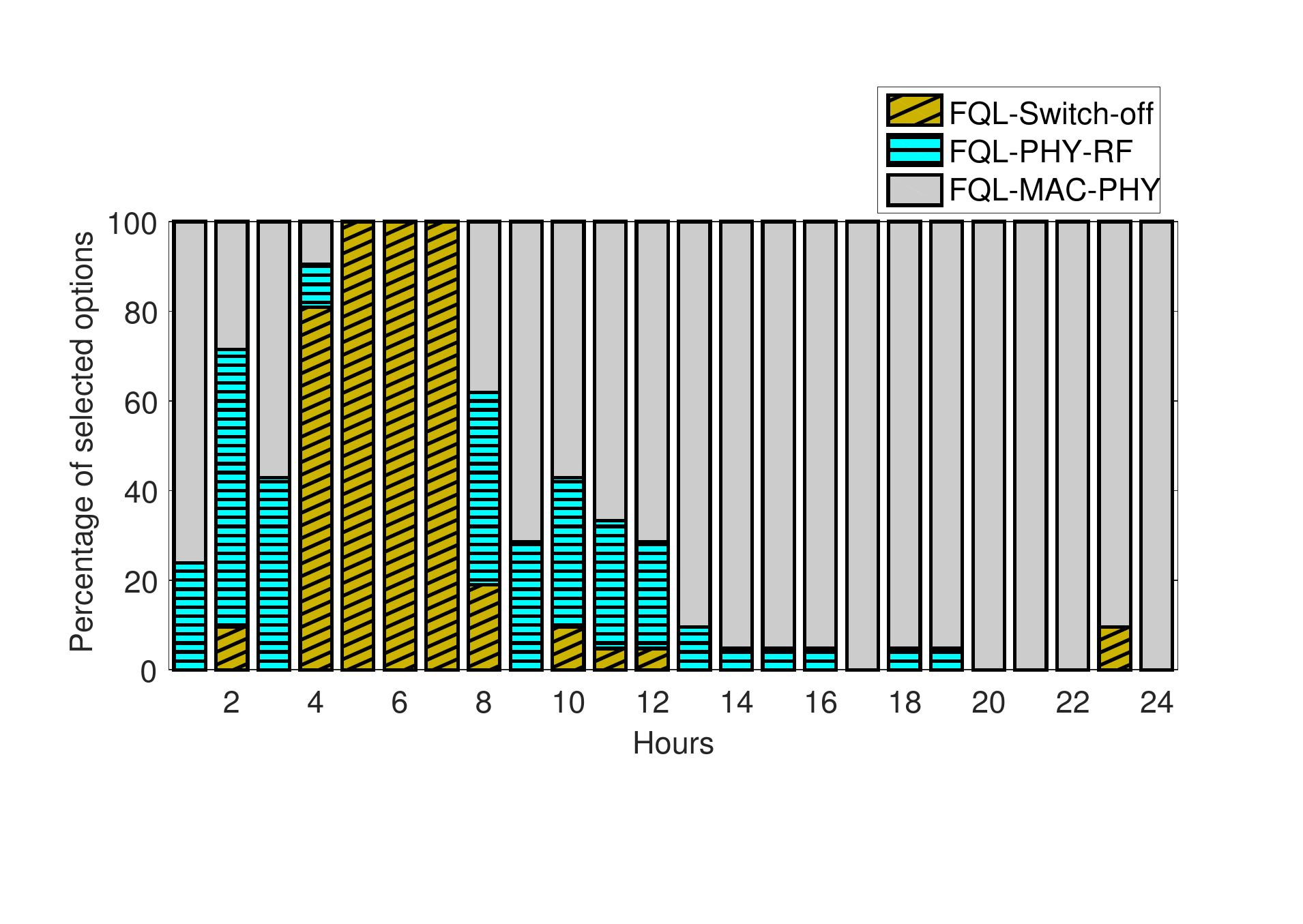}}\hspace{-2.2em}\quad
\subfloat[]{\includegraphics[width=2.35in]{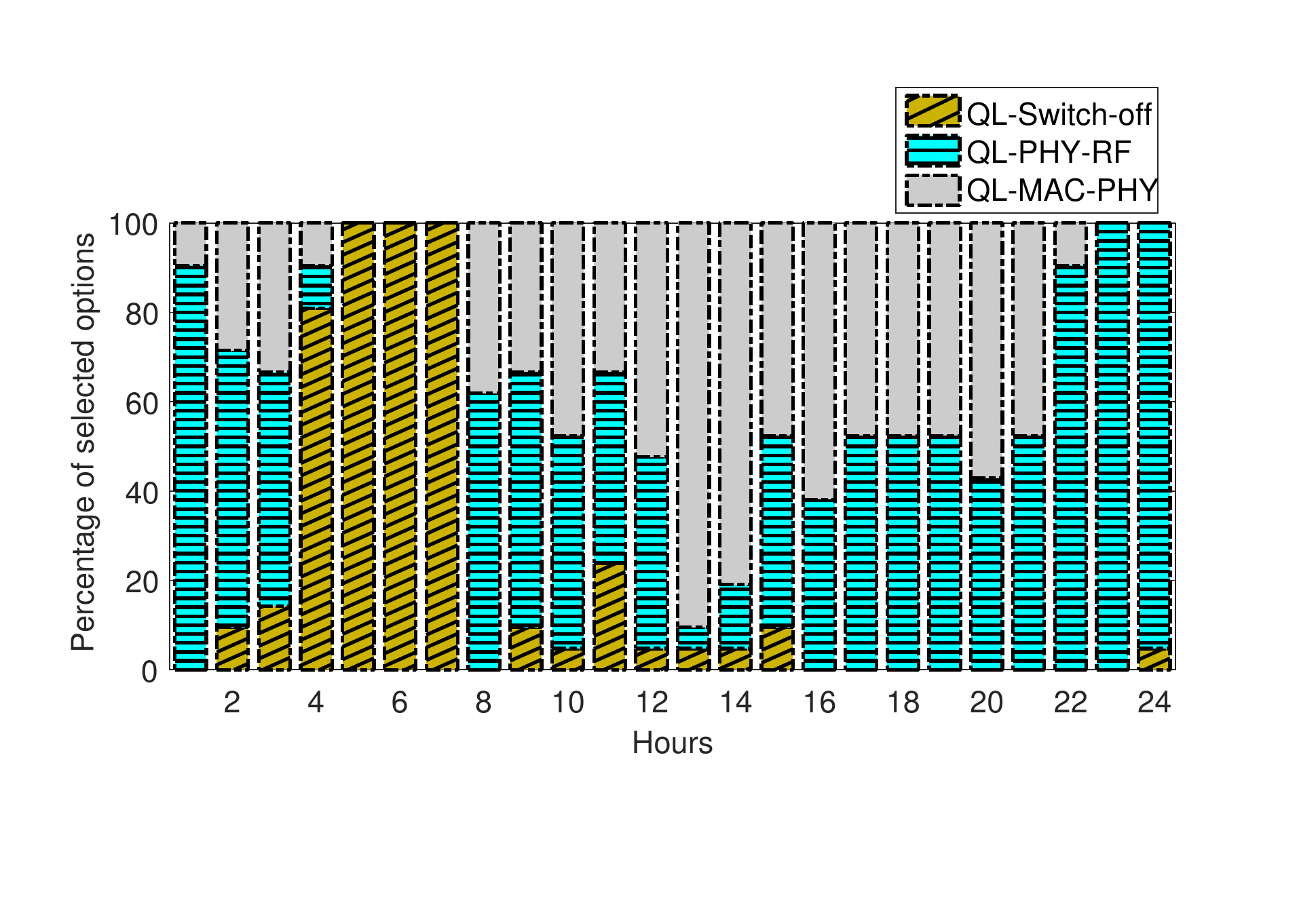}}\hspace{-2.0em}
 \vspace*{-0.6cm}
\caption{Average residential area summer day policy characteristics of: (a)~Off-line (b)~FQL (c)~QL }
\label{fig:summer-policy-residential}
\end{figure}

\begin{figure}[h]
\centering
\captionsetup{position=top}
\captionsetup[subfloat]{captionskip=-3pt}
 \hspace*{-0.5cm}\subfloat[]{\includegraphics[width=2.35in]{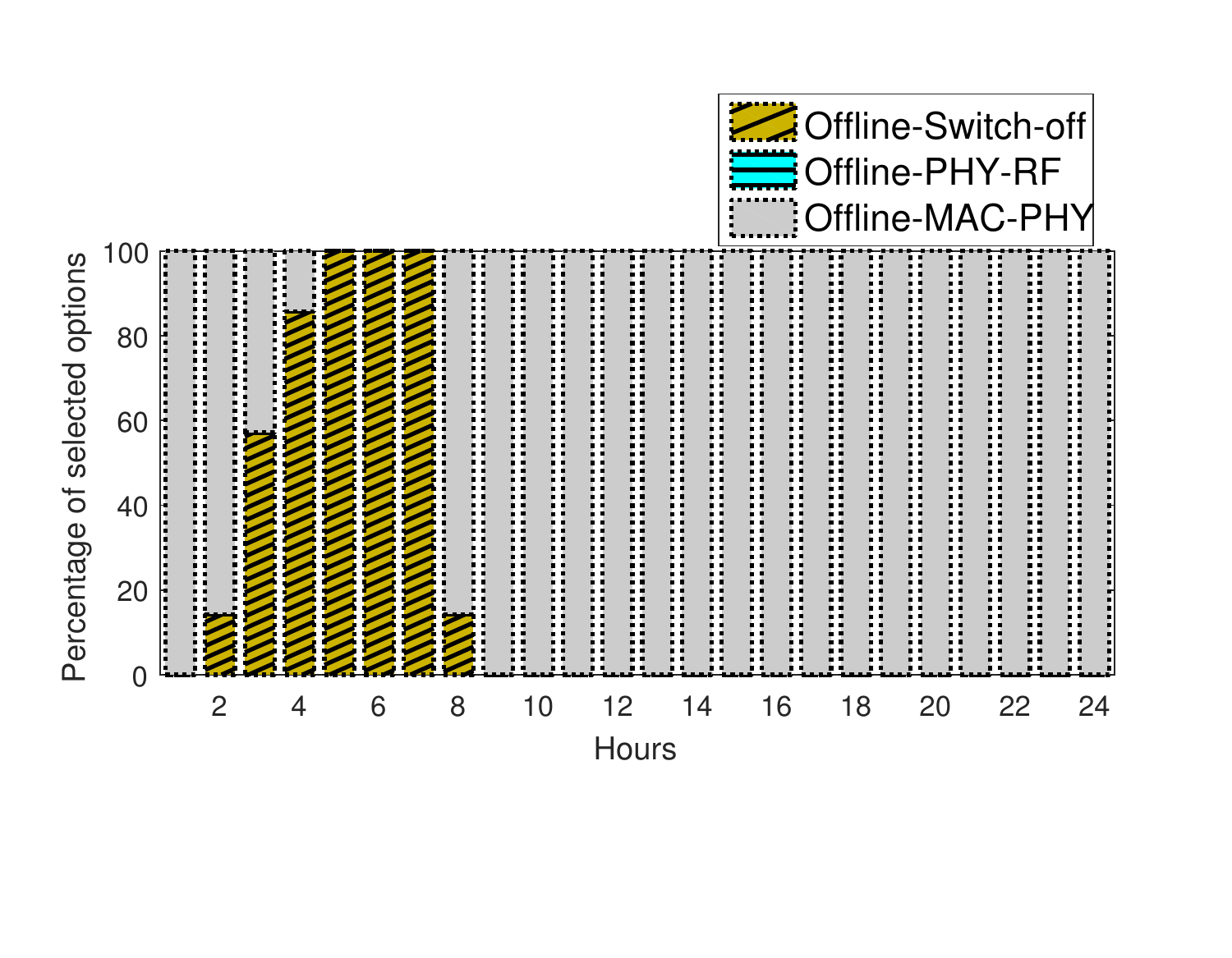}}\hspace{-2.2em}\quad
\subfloat[]{\includegraphics[width=2.35in]{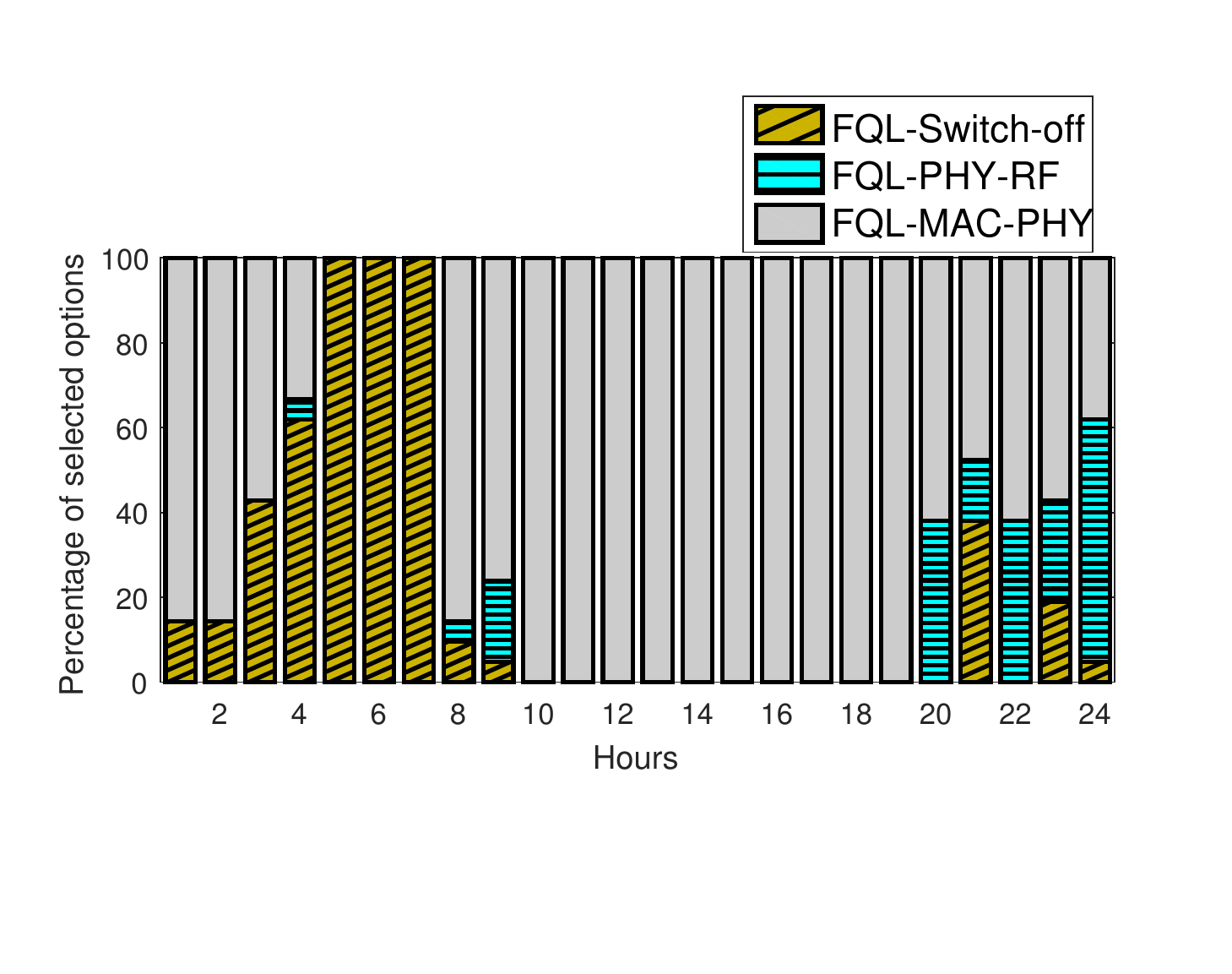}}\hspace{-2.2em}\quad
\subfloat[]{\includegraphics[width=2.35in]{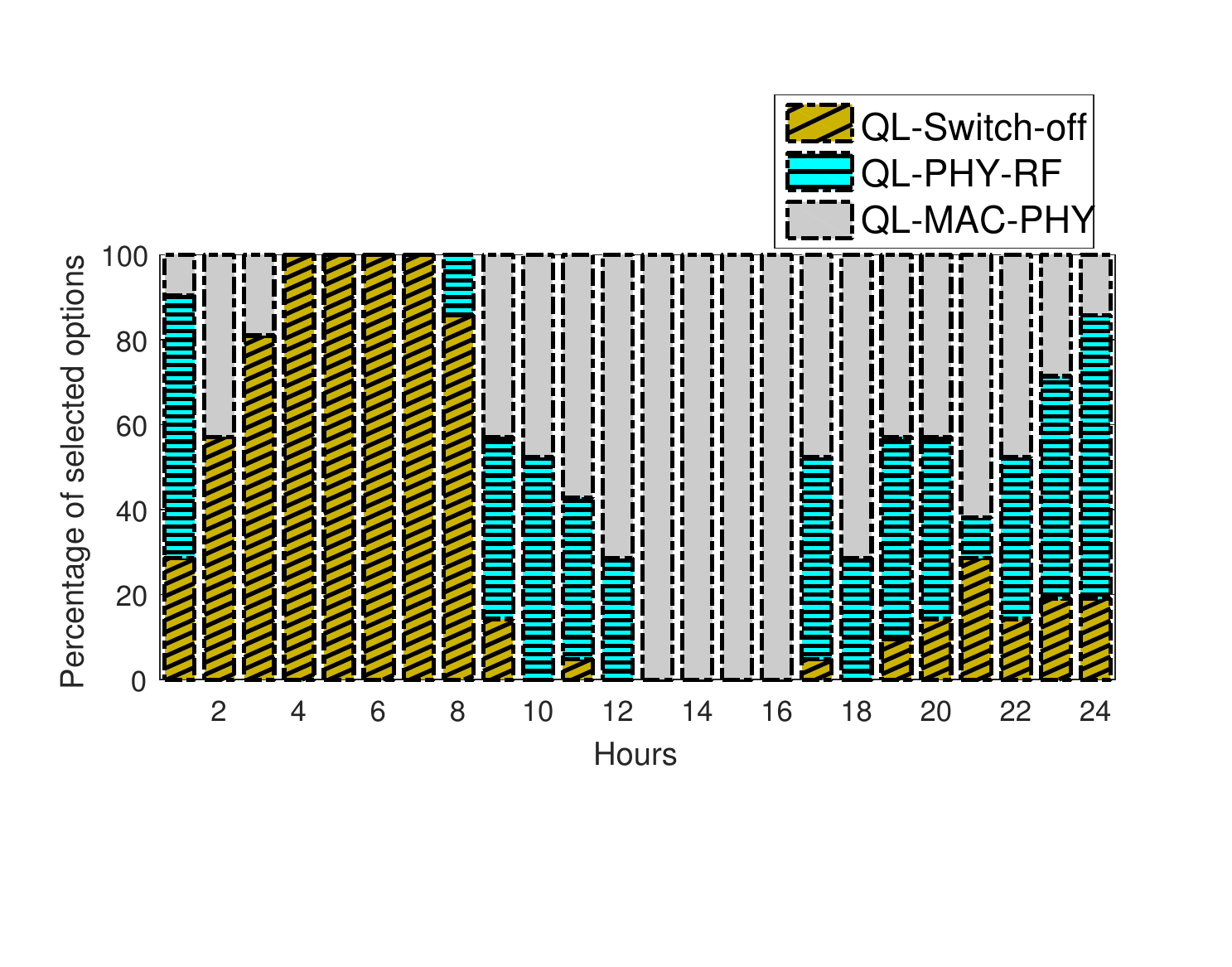}}\hspace{-2.0em}
 \vspace*{-0.6cm}
\caption{Average office area summer day policy characteristics of: (a)~Off-line (b)~FQL (c)~QL }
\label{fig:summer-policy-office}
\end{figure}

%[PD: PLEASE MERGE FIG10-11 IN THE SAME ROW AND CHANGE TEXT ACCORDINGLY.]
The policy behavior of both FQL and QL based controls for a system with $3$ vSCs are evaluated with respect to the off-line policy. The off-line solution, described in Section \ref{sec:network-model}, is based on DP and aimed at determining the performance bound of dynamic functional split placement when system dynamics information are known a-priori. The policy behaviors of an off-line, FQL and QL based controls for $3$ vSCs for an average winter day are depicted in Fig. \ref{fig:winter-policy-residential} and Fig. \ref{fig:winter-policy-office} for residential and office area, respectively. Moreover the functional split selection behaviors for an average summer day are shown in Fig. \ref{fig:summer-policy-residential} and Fig. \ref{fig:summer-policy-office} for off-line, FQL and QL controls in residential and office area, respectively. These results show that both FQL and QL polices usually switch-off most of the vSCs during very low traffic periods, as done by the off-line policy. However, the polices substantially differ in their respective functional split options selection when switched on. In this regard, QL is adopting a more conservative approach by selecting more PHY-RF split option as compared to the other solutions. In fact, FQL has a more similar behavior with respect to the off-line solution thanks to its higher flexibility in policy selection for both MAC-PHY and PHY-RF splits. In residential area and on average winter day, the MAC-PHY selection rates are $51\%$, $46\%$ and $23\%$ for off-line, FQL and QL controls respectively. For average summer day, the residential area MAC-PHY selection rate rises to $77\%$, $68\%$ and $34\%$ in off-line, FQL and QL solutions respectively. On the other hand, on average winter day in office area, the MAC-PHY selection rates are $62\%$, $50\%$ and $34\%$ for off-line, FQL and QL controls respectively. For average summer day, the office area MAC-PHY selection rate rises to $81\%$, $70\%$ and $44\%$ in off-line, FQL and QL solutions respectively. Hence, the FQL policy is able to have higher adaptation to the energy income of the seasons. It is also interesting to note that for an office area, the off-line solution configuration is predominantly between switch-off and MAC-PHY split. This can be clearly seen in Fig. \ref{fig:summer-policy-office},  where the off-line solution has $0\%$ PHY-RF selection rate on an average summer day.
% The figures (Fig. \ref{fig:switchoff-policy} and  Fig. \ref{fig:macphy-policy}) show than QL switching-off behavior is very close to the off-line policy where as FQL achieves closer MAC-PHY selection behavior than QL\emph{(MM: any idea on the reason behind this difference?)}
%\begin{figure*}
%\centering
%\hspace*{-3cm}\includegraphics[scale = 0.61]{winter_policy.eps}
%\vspace*{-3cm}
%\caption{Functional split options selected in an average winter day with $3$ vSCs in residential traffic profile} 
%\label{fig:winter-policy}
%\end{figure*}

%\begin{figure*}
%\centering
%\hspace*{-3cm}\includegraphics[scale = 0.61]{summer_policy.eps}
%\vspace*{-3cm}
%\caption{Functional split options selected in an average summer day with $3$ vSCs in residential traffic profile} 
%\label{fig:summer-policy}
%\end{figure*}

%[\vspace{-2cm}Off-line\label{winter-offline}\vspace*{-2cm}]
%[FQL\label{winter-fql}]
%[QL\label{winter-ql}

\subsubsection{Network Performance}

%The objective of the proposed QL and FQL based controls is to minimize the grid energy utilization while avoiding system outage. Hence, 
this section evaluates the performance of the polices obtained in the training phase in terms of grid energy consumption and system drop rate parameters for a year of operation. Hence, the same energy arrival and traffic demand profiles used in the training phase are used for the evaluation of the network performance. First we compare the performance of the policies against the off-line bound in a scenario with $3$ vSCs. Moreover, as a comparison bench mark, we considered uncoordinated solutions, i.e. U-FQL and U-QL presented in Section~\ref{ssec:u-control}, and a greedy approach. The Greedy (G) approach works by keeping the vSC in PHY-RF mode all the time as long as the level of the battery is above a certain threshold ($B_{th}$). We call this static configuration as G-PHY-RF. The network performance results are shown in Table \ref{tab:policy-comparison}. 
\begin{table}[H]
  \centering
\caption{Comparisons with respect to the off-line bound - $3$ vSCs}
\scalebox{1}{
\begin{tabular}{|c|c|c|c|c|}
\hline
\multirow{2}{*}{Algorithm} & \multicolumn{2}{|p{3cm}|}{\centering  Grid energy consumption (KWh)} & 
    \multicolumn{2}{|p{2cm}|}{\centering Average drop rate (\%)} \\
\cline{2-5}
 & Residential & Office& Residential & Office \\
\hline
Off-line&7403 &6744 & 0.3 & 0.02\\

FQL & 7695 (+3.9\%) & 7076 (+4.9\%)& 0.9 & 0.05\\

QL &  8150 (+10\%)& 7289 (+8.1\%) & 0.5 & 0.03 \\

U-FQL & 8000 (+8.0\%)& 7487 (+11\%) &  1.2 & 0.1\\

U-QL & 8202 (+10.8\%)&7688 (+14\%) &  0.7& 0.1\\

G-PHY-RF & 8320 (+11.2\%) & 8232 (+18.1\%) & 3.3 & 0.8\\

\hline
\end{tabular}}
\label{tab:policy-comparison}
\end{table}

These results show that FQL is able to achieve very low grid energy consumption which is only $4\%$ to $5\%$ higher than the energy consumption value obtained by the off-line bound for both office and residential profiles. On the other hand, QL policy consumes relatively higher grid energy which is about $8\%$ to $10\%$ higher than the off-line policy, for both office and residential area traffic. This can also be deduced from the policy behaviors in Fig. \ref{fig:winter-policy-residential}, Fig. \ref{fig:winter-policy-office}, Fig. \ref{fig:summer-policy-residential} and Fig. \ref{fig:summer-policy-office}, which show that FQL has higher MAC-PHY selection rate than QL. In particular, the FQL policy shows adaptation to a higher energy income in summer months by increasing the selection rate of MAC-PHY split. This behavior is also observed from the off-line solution. With higher MAC-PHY selection rate, more energy saving can be achieved since most of the BB processing functions are performed locally at vSCs. The results in Table \ref{tab:policy-comparison} also show that the policies without coordination, i.e. U-FQL and U-QL, exhibit lower performance than their coordinated counterparts, i.e. FQL and QL, respectively.

%[PD. PLEASE MERGE FIG16-17 AND FIG18-19. ALSO CHANGE THE TEXT ACCORDINGLY]

\begin{figure}[h]
\centering
\captionsetup{position=auto}
\captionsetup[subfloat]{captionskip=-1pt}
\vspace*{-0.0cm}
\subfloat[\label{energy-residential}]{\includegraphics[width=2.5in]{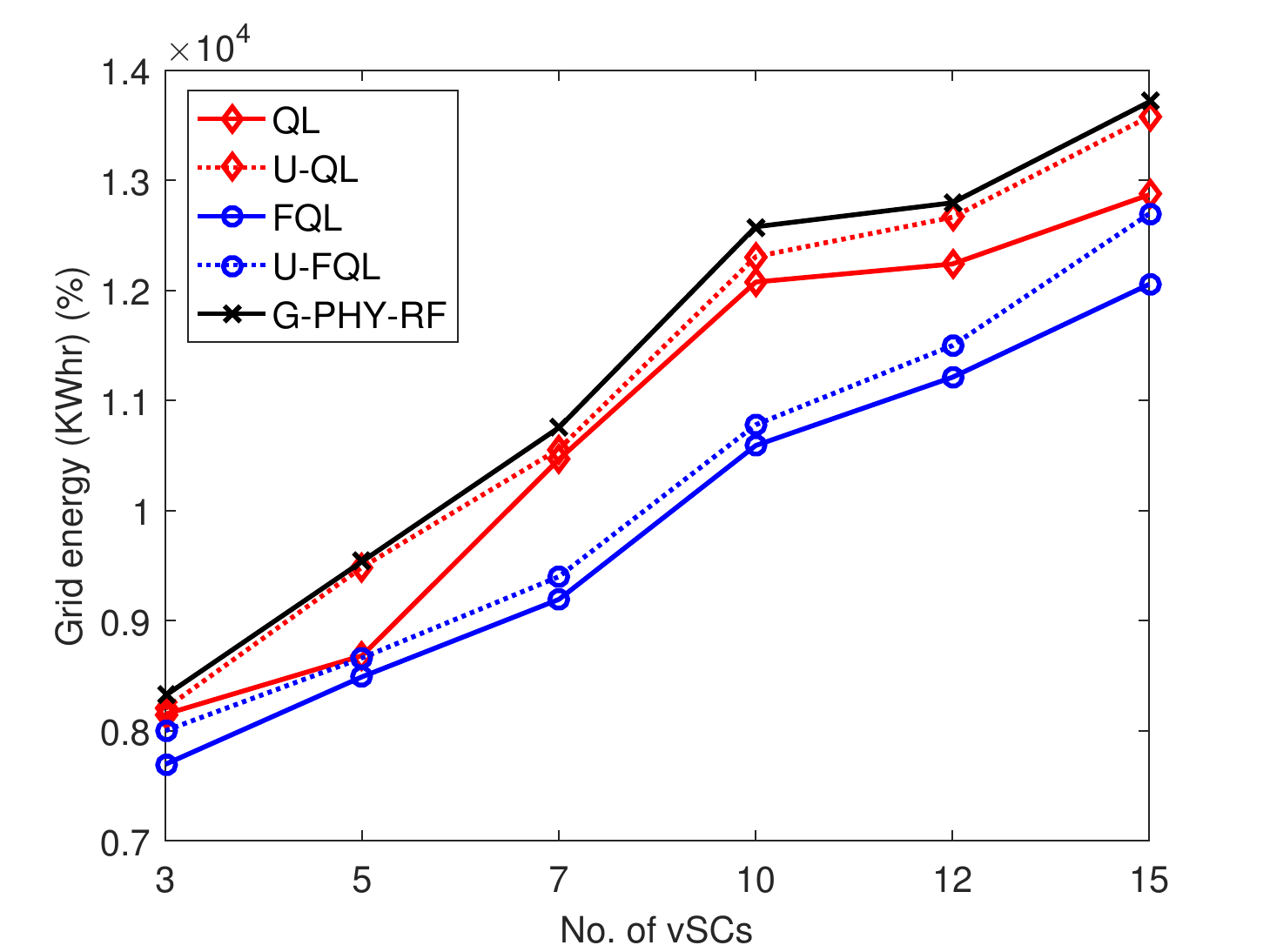}}\hspace{-0em}\quad
\subfloat[\label{energy-office}]{\includegraphics[width=2.5in]{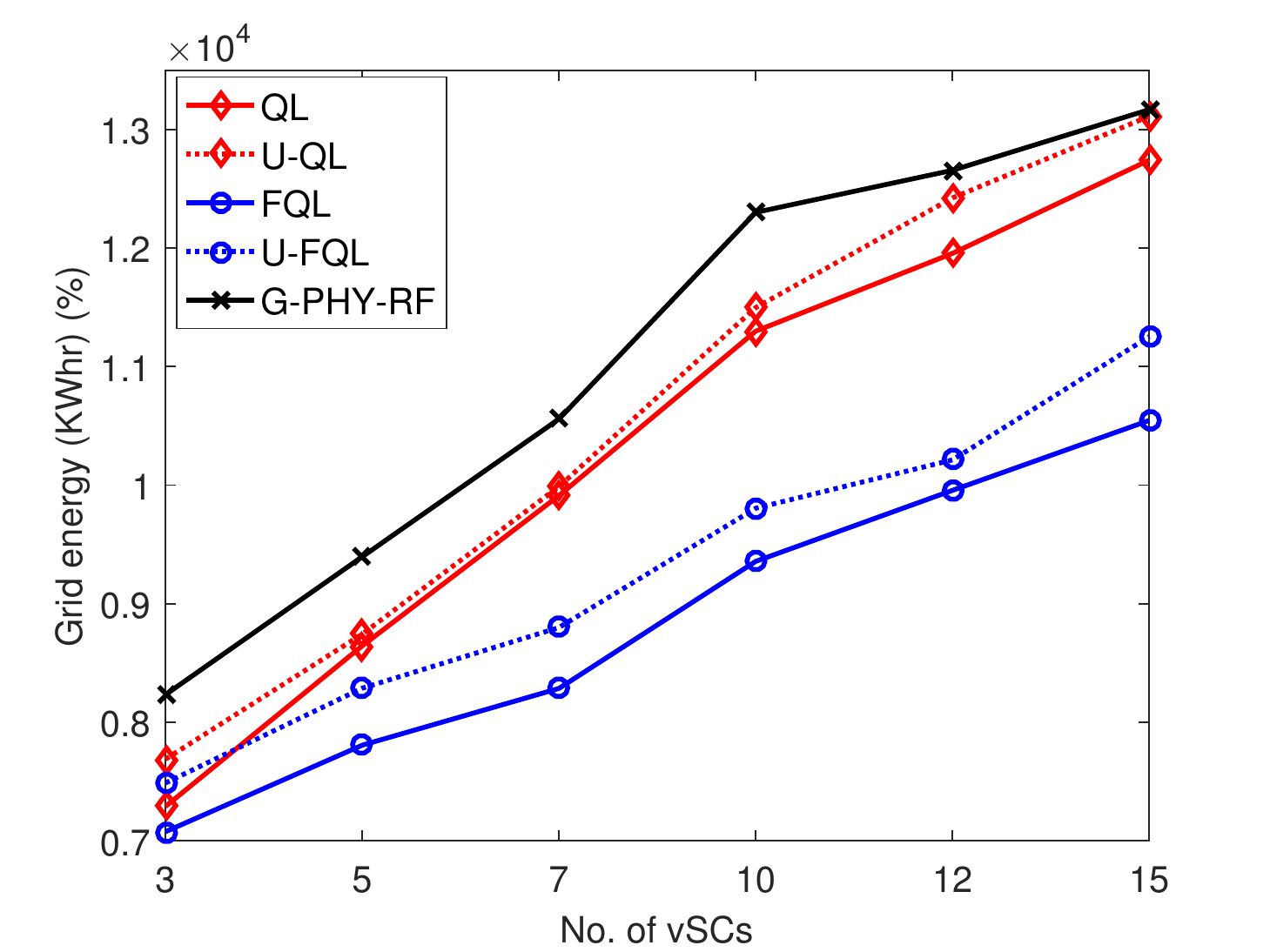}}\hspace{-0em}\quad
 \vspace*{-0.3cm}
\caption{Grid energy consumption comparison among FQL, QL, U-FQL, U-QL  and G-PHY-RF solutions: (a)~Residential (b)~Office }
\label{fig:energy-comparison}
\end{figure}

\begin{figure}[h]
\centering
\captionsetup{position=auto}
\captionsetup[subfloat]{captionskip=-1pt}
\vspace*{-0.0cm}
\subfloat[\label{drop-residential}]{\includegraphics[width=2.5in]{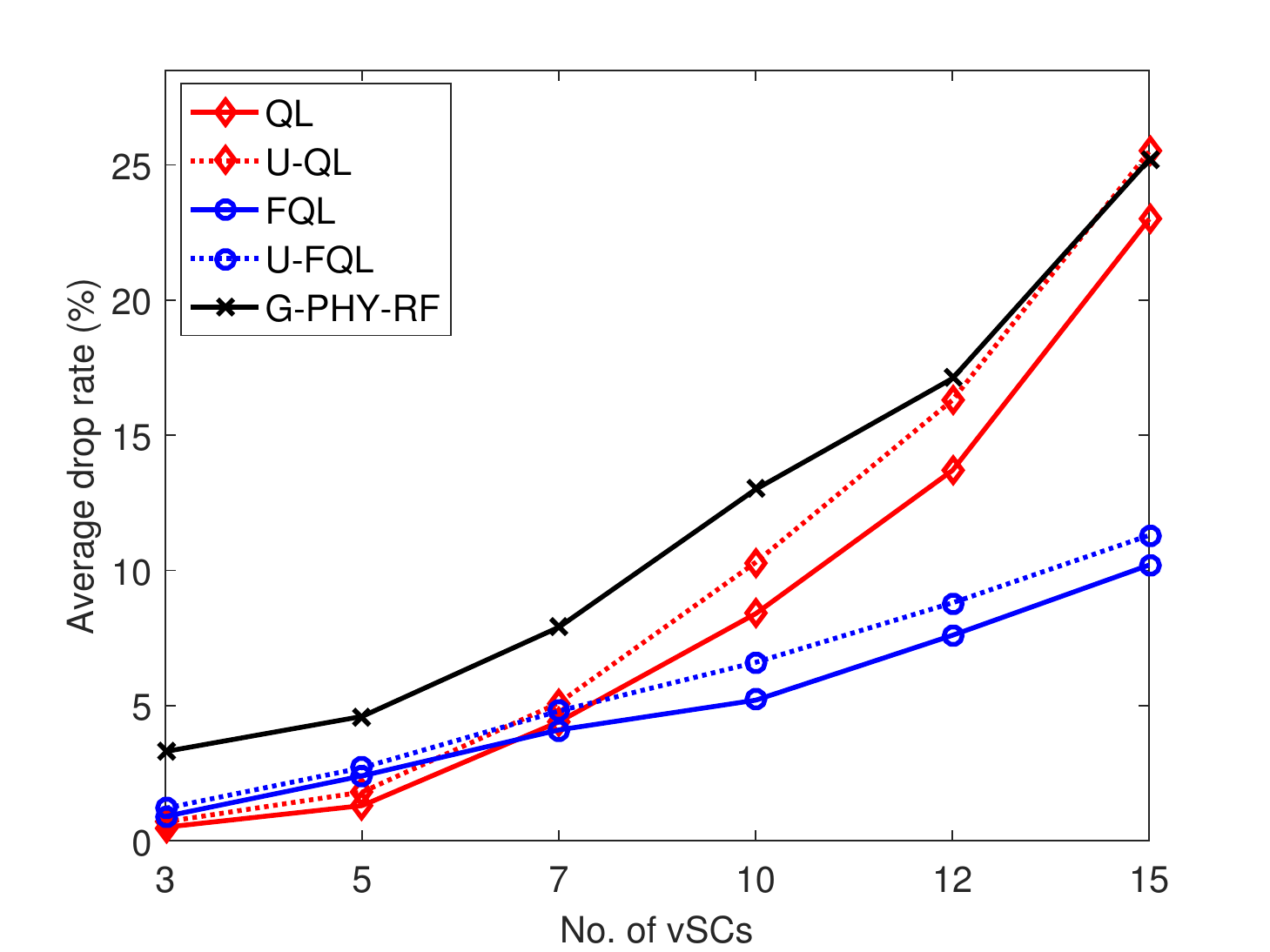}}\hspace{-0em}\quad
\subfloat[\label{drop-office}]{\includegraphics[width=2.5in]{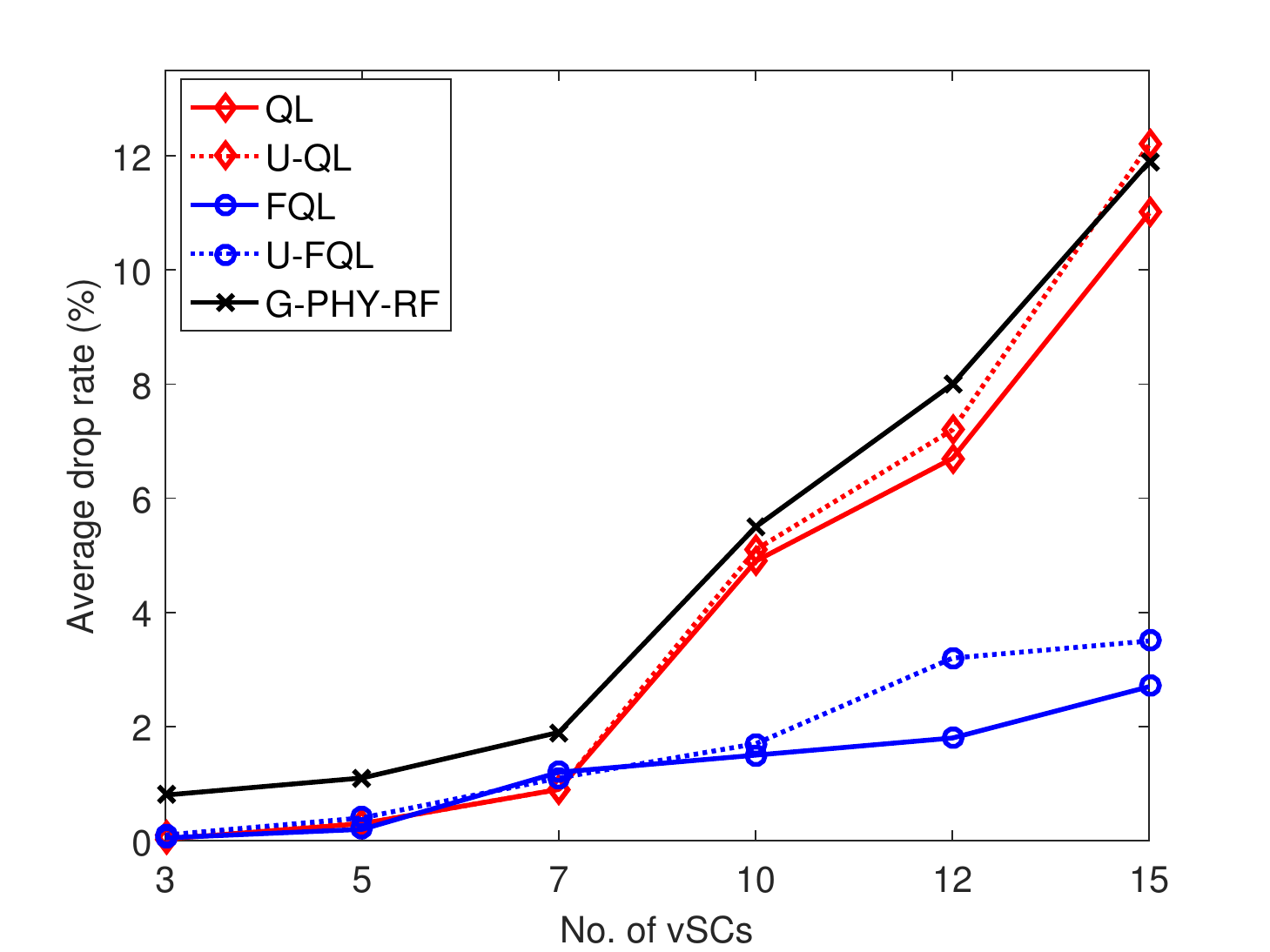}}\hspace{-0em}\quad
 \vspace*{-0.3cm}
\caption{Average drop rate comparison among FQL, QL, U-FQL, U-QL  and G-PHY-RF solutions: (a)~Residential (b)~Office }
\label{fig:drop-comparison}
\end{figure}

The grid energy consumption performances of FQL and QL based controllers in residential and office area for a year of operation for higher number of vSCs are shown in Fig. \ref{fig:energy-comparison}. Due to high computational demand of the off-line solution, we could not show the off-line bound results for vSCs higher than $3$. Moreover, the traffic drop rate performances of both FQL and QL controls in residential and office area scenario are shown in  Fig. \ref{fig:drop-comparison}. These results show that FQL policy performs better both in grid energy consumption and average drop rate in both residential and office profiles for higher number of vSCs. The performance gap between FQL and the other solutions is growing with the number of vSCs ($7$, $10$, $12$ and $15$ vSCs). In residential area traffic, the FQL controller achieves an energy saving of up to $12\%$ and an average drop rate of up to $10\%$ less than the QL control. Moreover, FQL controller is able to achieve energy saving of up to $17\%$ and average drop rate of up to $8\%$ less than QL control in an office area traffic profile. The better performance by the FQL is aligned with the higher cumulative rewards obtained by the FQL controller as shown in Fig. \ref{fig:cr-bar}. In addition, the results of uncoordinated solutions,  i.e. U-FQL and U-QL, are shown for comparison. The solutions without MBS traffic load information, i.e. U-QL and U-FQL, have lower performances than the proposed QL and FQL counterparts both in energy consumption and average drop rate. In addition, the static configuration policy i.e. G-PHY-RF, presents the lowest performance, below than the RL based methods in both grid energy saving and average drop rate.

\subsection{Policy Validation}
\label{sec:validation}

In this section we evaluate the behavior of the system in real deployment scenario with a training performed off-line with simulation. In detail, we will validate the proposed FQL and QL based controllers using a new environment, which is characterized by an energy arrival and traffic demand profiles which are different from the environment used for simulated training. In this case we are using the algorithms with pre-trained Q-values and with an exploration rate of  $5\%$. 
%Hence, the vSCs are taking actions with maximum Q-values with $95\%$ of the time. 
The validation of the policies along with the training environment policy evaluation for $3$, $5$, $7$, $10$, $12$ and $15$ vSCs for a year of operation are shown in Table \ref{tab:policy-validation}. The validation results in Table \ref{tab:policy-validation} show that both FQL and QL are able to adapt their behaviors to the new validation environment. This is confirmed by both grid energy and average drop rate performances that are very close to the corresponding policy evaluation results.
%\emph{[MM: maybe we can report the  results of the training for sake of readability]}. 
These results give an insight that using simulated trained Q-values / rules-actions consequents for QL / FQL respectively, continuously exploring new actions in the new environment and updating the corresponding Q-tables is a viable approach in real deployment scenarios. 
\begin{table*}[ht]
  \centering
\captionsetup{justification=centering}
\caption{Policy validation results\\
{\small (R-T: Residential Training, O-T: Office Training, R-V: Residential Validation, O-V:  Office Validation)}}
\scalebox{1}{
\begin{tabular}{|c|c|c|c|c|c|c|c|c|c|}
\hline
{\centering No. of vSCs} & \multirow{2}{*}{\centering Algorithm} & \multicolumn{4}{|p{4cm}|}{\centering  Grid energy consumption (KWh)} & 
    \multicolumn{4}{|p{4cm}|}{\centering Average drop rate (\%)} \\
\cline{3-10}
 && R - T & O - T & R - V & O - V & R - T & O - T & R - V & O - V\\
\hline
3&FQL &  7695  & 7076 & 7757 & 7211& 0.9 & 0.05 & 0.8 & 0.02\\
&QL &  8150 & 7289 & 8330 & 7495 & 0.5 & 0.03 & 0.6 & 0.03\\
\hline
5&FQL & 8490 & 7806 & 8527 & 7805 & 2.4 &0.2 & 3& 0.18\\
&QL & 8682 & 8644 & 8903 & 8639 &1.3 & 0.3&1.4&0.35\\
\hline
7&FQL & 9193 & 8285 &  9189 & 8330& 4.1 & 1.2 &4.8&1.37\\
&QL & 10466 & 9912 & 10482 & 9885 & 4.4 & 0.9 &4.7 & 1.1\\
\hline
10&FQL & 10591 & 9357&  10606 & 9367 & 5.2 &1.5 & 5.4& 1.5\\
&QL & 12076& 11299 & 12177 & 11291 & 8.4 & 4.9 & 8.8 & 5.1\\
\hline
12&FQL & 11211 & 9956 &  11222 & 10019 & 7.6 & 1.8 & 8.4 & 2.0\\
&QL & 12240 & 11955 & 12216 & 11967 & 13.7 & 6.7 & 14. 1 & 6.7\\
\hline
15&FQL & 12056 & 10546 &  12182 & 10623 & 10.2 & 2.7 & 10.6 & 3.2\\
&QL & 12869 & 12742 & 12917 & 12711 & 23 & 11 & 23.4 & 11.1\\
\hline
\end{tabular}}
\label{tab:policy-validation}
\end{table*}

\begin{figure}
\centering
\includegraphics[scale = 0.6]{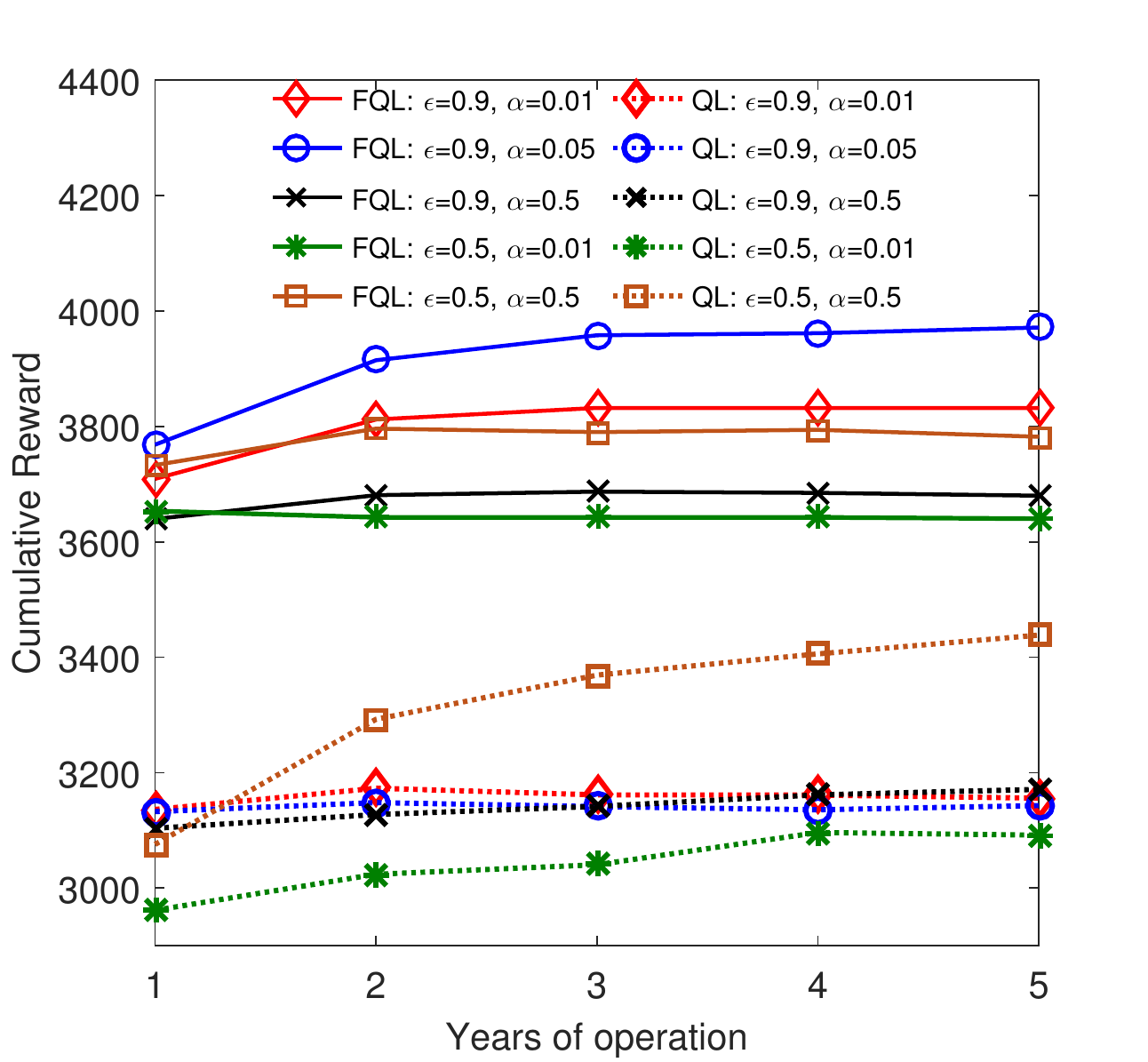}
\caption{Cumulative reward for run-time training of FQL and QL in residential area for $3$ vSCs} 
\label{fig:runtime-cr}
\end{figure}

\begin{figure}
\centering
\includegraphics[scale = 0.6]{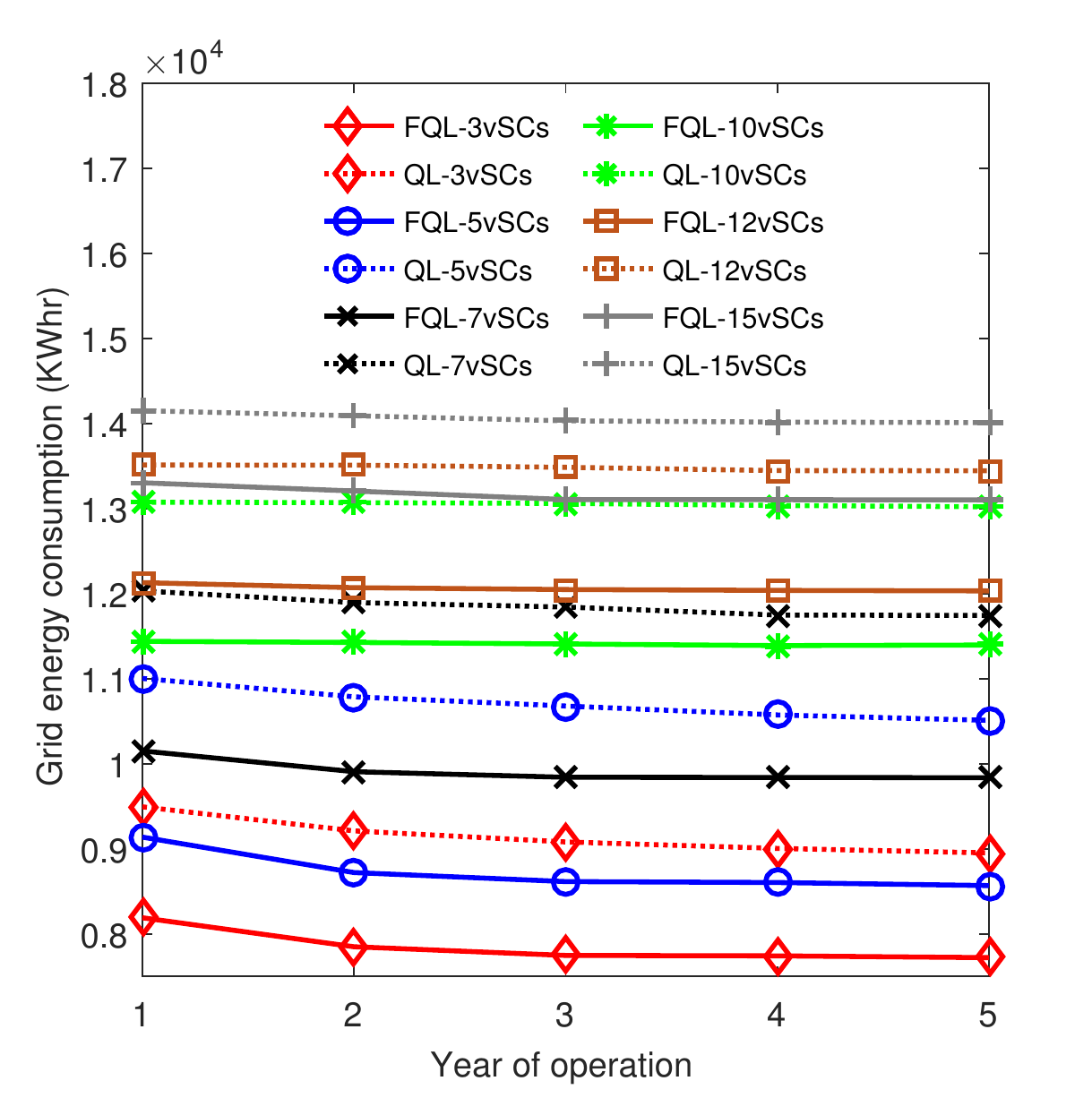}
\caption{Grid energy consumption of run-time FQL and QL controls in residential scenario for $3$, $5$, $7$, $10$, $12$ and $15$ vSCs} 
\label{fig:runtime-energy}
\end{figure}

\begin{figure}
\centering
\includegraphics[scale = 0.6]{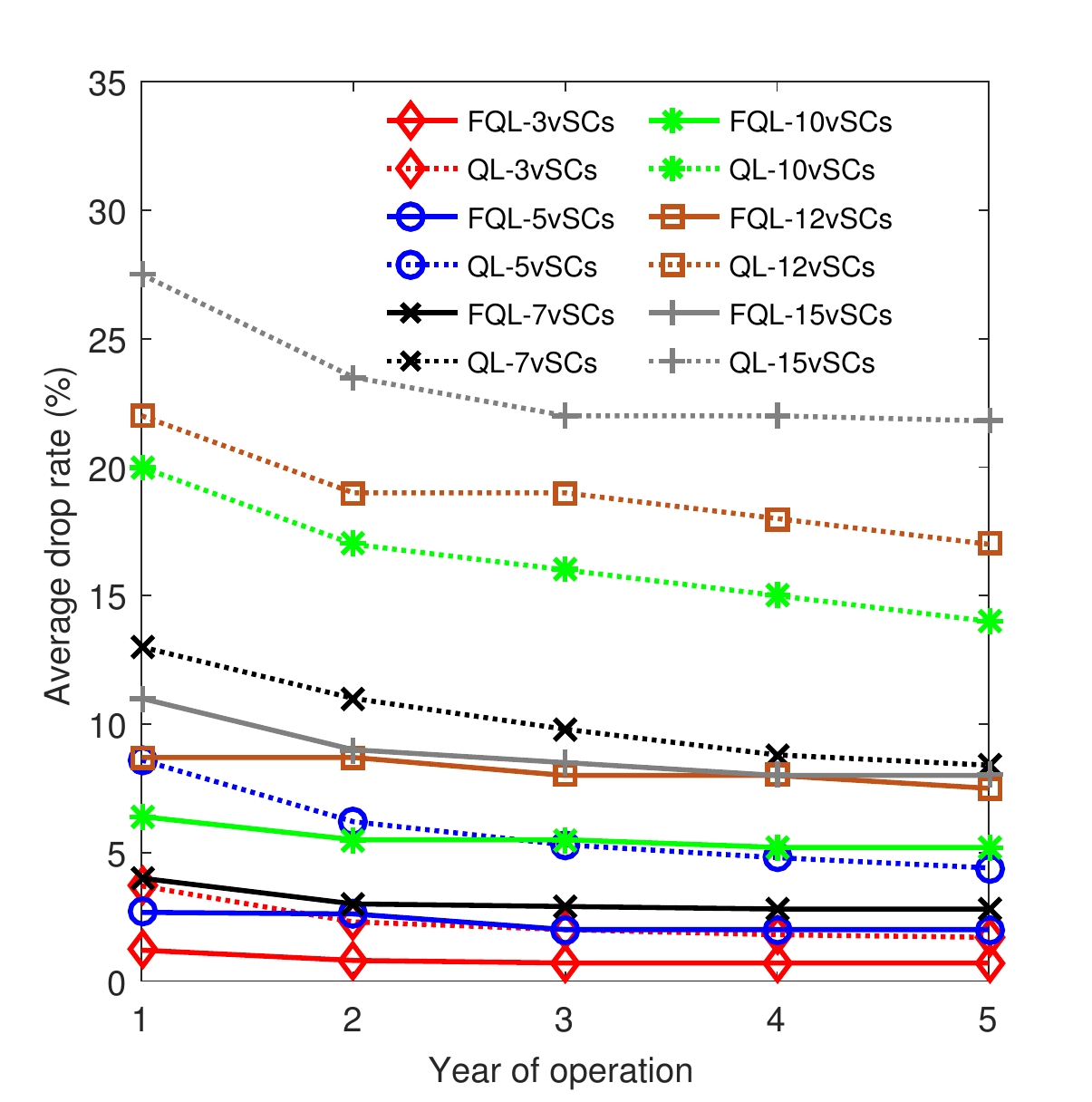}
\caption{Average drop rate of run-time FQL and QL controls in residential scenario for $3$, $5$, $7$, $10$, $12$ and $15$ vSCs} 
\label{fig:runtime-drop}
\end{figure}

\subsection{Run-time training}

Here, we perform the evaluation of the proposed FQL and QL based controls in run-time deployment scenario without pre-training, i.e., the vSCs are learning on the job while they are in operation. In this case, all the Q-values / rules-actions consequences are initialized to $0$ for QL and FQL, respectively. An exploration/exploitation strategy is used for the learning of vSCs. In order to determine the effect of the learning parameters on run-time FQL and QL performances, we have compared the on-line training for different sets of learning rate ($\alpha$) and exploration ($\epsilon$) parameters for $3$ vSCs. These results are shown in Fig. \ref{fig:runtime-cr}. The results show that FQL is able to gain higher cumulative rewards than QL starting from the first year of operation.
%[PD: WHAT DO YOU MEAN FOR ACCUMULATE REWORDS? IT IS ABOUT THE STABILITY/CONVERGENCE? IF SO SHOULDN'T WE WRITE SOMENTHING ON THE STABILIZATION OF THE DIFFERENT VALUES OF Q IN THE TABLE?]. 
As a result, it is more suitable for run-time training of the vSCs than QL. Moreover, lower values of the learning rate are better for FQL, whereas QL requires relatively higher learning rate. The exploration rate parameter shown in  Fig. \ref{fig:runtime-cr} are initial exploration rates, which are continuously discounted as the training progresses until reaching the minimum level of exploration, which is set at $5\%$. 

The grid energy consumption performances of both FQL and QL controls for a run-time training and operation are shown in Fig. \ref{fig:runtime-energy}. Moreover the average drop rate performances of run-time controls are shown in  Fig. \ref{fig:runtime-drop}.
These results show that FQL policy is more suitable for run-time application, as shown both in terms of grid energy consumption and average drop rate.
% The FQL is able to learn faster and can achieve better cumulative reward starting from first year of operation. On the other hand, QL is learning slowly towards minimizing both grid energy consumption and drop rate.
Run-time FQL is able to gain an energy saving ranging from $10\%$ to $17\%$ with an average drop rate of $2.5\%$ to $13\%$ less, than run-time QL, in the first year of deployment.

 However, compared to policy validation results based on pre-trained agents, shown in Table \ref{tab:policy-validation}, the run-time training and operation results shown in Fig. \ref{fig:runtime-energy} and Fig. \ref{fig:runtime-drop} display lower performances. For FQL controller, the validation results of pre-trained agents shown in Table \ref{tab:policy-validation}, achieve energy saving ranging from $5\%$ to $9.5\%$ with a drop rate of $0.4\%$ to $1\%$  less, than the run-time results in  Fig. \ref{fig:runtime-energy} and Fig. \ref{fig:runtime-drop}. Moreover, for QL, the validation results in Table \ref{tab:policy-validation}, show an energy saving ranging from $7\%$  to $19\%$ with a drop rate of $3\%$ to $12\%$ less, than the run-time results. 
%[PD. PLEASE CHECK THESE SENTENCES. IT SEEMS THAT ENERGY SAVINGS ARE LESS IN PRE-TRAINED SCENARIO. IT IS THE CASE OF DROP, BUT NOT FOR ENERGY, RIGHT? MAYBE IT IS JUST THE CASE TO WRITE THAT ENERGY SAVINGS ARE HIGHER AND DROP LOWER IN THE PRE-TRAINED CASE]
Hence, to get closer to the optimization goals, it is better to initialize vSCs' agents with some knowledge prior to their deployment. This can be in the form of simulated training of the vSCs, as shown in \ref{sec:offline-training}. As a result, training of the vSCs in a simulative environment prior to their deployment and allowing them to explore new knowledge while in operation is a more appropriate approach in real deployments.
%\section{Open Issues}
%This section highlights some of the open issues regarding the control of energy harvesting small cells in MEC scenarios that require further research.\\

\section{Conclusions}
\label{sec:conclusions} 

In this paper we have proposed a network scenario where the computational processes of the vSCs powered solely by energy harvesters and batteries may be shared with on grid-connected central MEC-server at the MBS for part of their BB processing. We have stated the energy minimization problem and proposed multi-agent RL to solve it. Distributed Fuzzy Q-Learning and Q-Learning  on-line algorithms are tailored for our purposes. Coordination among the multiple agents is achieved by broadcasting system level information (i.e. the traffic load at MBS) to the independent learners. Finally, we have evaluated the network performance (in terms of energy consumption and traffic drop rate) by our multi-agent RL solutions with different levels of coordination and compared them against the off-line performance bound and static solutions. 

Our results confirm that coordination via broadcasting may achieve higher system level gains than un-coordinated solutions and cumulative rewards closer to the off-line bounds. Moreover, our analysis permits to evaluate the benefits of continuous state/action representation for the learning algorithms in terms of faster convergence, higher cumulative reward and more adaptation to changing environments.


%%%%%%%%%%%%%%%%%%%%%%%%%%%%%%%%%%%%%%%%%%%%%%%%%%%%%%%%%%%%%%%%%%%%%%%%%%%%%%%%

%%%%%%%%%%%%%%%%%%%%%%%%%%%%%%%%%%%%%%%%%%%%%%%%%%%%%%%%%%%%%%%%%%%%%%%%%%%%%%%%

%%%%%%%%%%%%%%%%%%%%%%%%%%%%%%%%%%%%%%%%%%%%%%%%%%%%%%%%%%%%%%%%%%%%%%%%%%%%%%%%
\section*{Acknowledgement}
This work has received funding from the European Union's Horizon 2020 research and innovation programme under the Marie \mbox{Sklodowska-Curie} grant agreement No 675891 \mbox{(SCAVENGE)} and by Spanish MINECO grant TEC2017-88373-R (5G-REFINE).

\bibliographystyle{IEEEtran}
\bibliography{ref}
\enlargethispage{50mm}
%%%%%%%%%%%%%%%%%%%%%%%%%%%%%%%%%%%%%%%%%%%%%%%%%%%%%%%%%%%%%%%%%%%%%%%%%%%%%%%%
\end{document}